\pdfoutput=1
\documentclass[12pt,a4paper]{article}
\usepackage{ifthen} \newboolean{pdflatex}
\setboolean{pdflatex}{true} 
\newboolean{articletitles}
\setboolean{articletitles}{true} 
\newboolean{uprightparticles}
\setboolean{uprightparticles}{false} 
\newboolean{inbibliography}
\setboolean{inbibliography}{false} 
\usepackage[top=1in, bottom=1.25in, left=1in, right=1in]{geometry}

\usepackage{microtype}
\usepackage{lineno}  % for line numbering during review
\usepackage{xspace} % To avoid problems with missing or double spaces after
\usepackage{caption} %these three command get the figure and table captions automatically small

\usepackage{graphicx}  % to include figures (can also use other packages)
\usepackage{color}
\usepackage{colortbl}
\graphicspath{{./figs/}} % Make Latex search fig subdir for figures

\usepackage{amsmath} % Adds a large collection of math symbols
\usepackage{amssymb}
\usepackage{amsfonts}
\usepackage{upgreek} % Adds in support for greek letters in roman typeset

\newcommand*\patchAmsMathEnvironmentForLineno[1]{%
\expandafter\let\csname old#1\expandafter\endcsname\csname #1\endcsname
\expandafter\let\csname oldend#1\expandafter\endcsname\csname
end#1\endcsname
 \renewenvironment{#1}%
   {\linenomath\csname old#1\endcsname}%
   {\csname oldend#1\endcsname\endlinenomath}%
}
\newcommand*\patchBothAmsMathEnvironmentsForLineno[1]{%
  \patchAmsMathEnvironmentForLineno{#1}%
  \patchAmsMathEnvironmentForLineno{#1*}%
}
\AtBeginDocument{%
\patchBothAmsMathEnvironmentsForLineno{equation}%
\patchBothAmsMathEnvironmentsForLineno{align}%
\patchBothAmsMathEnvironmentsForLineno{flalign}%
\patchBothAmsMathEnvironmentsForLineno{alignat}%
\patchBothAmsMathEnvironmentsForLineno{gather}%
\patchBothAmsMathEnvironmentsForLineno{multline}%
\patchBothAmsMathEnvironmentsForLineno{eqnarray}%
}

\usepackage{hyperxmp}

\usepackage{hyperref}    % Hyperlinks in references

\usepackage[all]{hypcap} % Internal hyperlinks to floats.

\usepackage{xspace} 
\usepackage{upgreek}

\def\lhcb {\mbox{LHCb}\xspace}

\def\MagUp {\mbox{\em Mag\kern -0.05em Up}\xspace}

\ifthenelse{\boolean{uprightparticles}}%
{

 \def\Pmu         {\ensuremath{\upmu}\xspace}                 
 \def\Pnu         {\ensuremath{\upnu}\xspace}                 
                  
 \def\Ppi         {\ensuremath{\uppi}\xspace}

 \def\PDelta      {\ensuremath{\Delta}\xspace}                 
 \def\PXi      {\ensuremath{\Xi}\xspace}                 
 \def\PLambda      {\ensuremath{\Lambda}\xspace}                 
 \def\PSigma      {\ensuremath{\Sigma}\xspace}                 
 \def\POmega      {\ensuremath{\Omega}\xspace}                 
 \def\PUpsilon      {\ensuremath{\Upsilon}\xspace}                 
 
 \def\PB      {\ensuremath{\mathrm{B}}\xspace}                 
                  
 \def\PD      {\ensuremath{\mathrm{D}}\xspace}

 \def\PK      {\ensuremath{\mathrm{K}}\xspace}

 \def\Pb      {\ensuremath{\mathrm{b}}\xspace}                 
 \def\Pc      {\ensuremath{\mathrm{c}}\xspace}                 
 \def\Pd      {\ensuremath{\mathrm{d}}\xspace}                 
 \def\Pe      {\ensuremath{\mathrm{e}}\xspace}

 \def\Pi      {\ensuremath{\mathrm{i}}\xspace}

 \def\Pu      {\ensuremath{\mathrm{u}}\xspace}

}
{

 \def\Pmu         {\ensuremath{\mu}\xspace}                 
 \def\Pnu         {\ensuremath{\nu}\xspace}                 
                  
 \def\Ppi         {\ensuremath{\pi}\xspace}

 \mathchardef\PDelta="7101
 \mathchardef\PXi="7104
 \mathchardef\PLambda="7103
 \mathchardef\PSigma="7106
 \mathchardef\POmega="710A
 \mathchardef\PUpsilon="7107
                  
 \def\PB      {\ensuremath{B}\xspace}                 
                  
 \def\PD      {\ensuremath{D}\xspace}

 \def\PK      {\ensuremath{K}\xspace}

 \def\Pb      {\ensuremath{b}\xspace}                 
 \def\Pc      {\ensuremath{c}\xspace}                 
 \def\Pd      {\ensuremath{d}\xspace}                 
 \def\Pe      {\ensuremath{e}\xspace}

 \def\Pi      {\ensuremath{i}\xspace}

 \def\Pu      {\ensuremath{u}\xspace}

}

\makeatletter
\ifcase \@ptsize \relax% 10pt
  \newcommand{\miniscule}{\@setfontsize\miniscule{4}{5}}% \tiny: 5/6
\or% 11pt
  \newcommand{\miniscule}{\@setfontsize\miniscule{5}{6}}% \tiny: 6/7
\or% 12pt
  \newcommand{\miniscule}{\@setfontsize\miniscule{5}{6}}% \tiny: 6/7
\fi
\makeatother

\DeclareRobustCommand{\optbar}[1]{\shortstack{{\miniscule (\rule[.5ex]{1.25em}{.18mm})}
  \\ [-.7ex] $#1$}}

\def\en         {{\ensuremath{\Pe^-}}\xspace}   % electron negative (\em is taken)
\def\ep         {{\ensuremath{\Pe^+}}\xspace}

\def\mun        {{\ensuremath{\Pmu^-}}\xspace} % muon negative (\mum is taken)

\def\neub       {{\ensuremath{\overline{\Pnu}}}\xspace}

\def\neumb      {{\ensuremath{\neub_\mu}}\xspace}
\def\uquark    {{\ensuremath{\Pu}}\xspace}

\def\dquark    {{\ensuremath{\Pd}}\xspace}

\def\cquark    {{\ensuremath{\Pc}}\xspace}

\def\bquark    {{\ensuremath{\Pb}}\xspace}

\def\pion   {{\ensuremath{\Ppi}}\xspace}
\def\piz    {{\ensuremath{\pion^0}}\xspace}

\def\pip    {{\ensuremath{\pion^+}}\xspace}
\def\pim    {{\ensuremath{\pion^-}}\xspace}

\def\kaon    {{\ensuremath{\PK}}\xspace}
  \def\Kbar    {{\kern 0.2em\overline{\kern -0.2em \PK}{}}\xspace}

\def\KorKbar    {\kern 0.18em\optbar{\kern -0.18em K}{}\xspace}

\def\Km      {{\ensuremath{\kaon^-}}\xspace}

  \def\Dbar    {{\kern 0.2em\overline{\kern -0.2em \PD}{}}\xspace}
\def\D       {{\ensuremath{\PD}}\xspace}

\def\DorDbar    {\kern 0.18em\optbar{\kern -0.18em D}{}\xspace}

\def\Dstarp  {{\ensuremath{\D^{*+}}}\xspace}

\def\Bbar    {{\ensuremath{\kern 0.18em\overline{\kern -0.18em \PB}{}}}\xspace}

\def\BorBbar    {\kern 0.18em\optbar{\kern -0.18em B}{}\xspace}

\def\Bzb     {{\ensuremath{\Bbar{}^0}}\xspace}

  \def\Y#1S{\ensuremath{\PUpsilon{(#1S)}}\xspace}% no space before {...}!

\def\Xires       {{\ensuremath{\PXi}}\xspace}

\def\Lz          {{\ensuremath{\PLambda}}\xspace}
\def\Lbar        {{\ensuremath{\kern 0.1em\overline{\kern -0.1em\PLambda}}}\xspace}
\def\LorLbar    {\kern 0.18em\optbar{\kern -0.18em \PLambda}{}\xspace}
\def\Lambdares   {{\ensuremath{\PLambda}}\xspace}

\def\Sigmares    {{\ensuremath{\PSigma}}\xspace}

\def\Lb      {{\ensuremath{\Lz^0_\bquark}}\xspace}

\def\Lc      {{\ensuremath{\Lz^+_\cquark}}\xspace}

\def\Xib     {{\ensuremath{\Xires_\bquark}}\xspace}

\newcommand{\decay}[2]{\ensuremath{#1\!\to #2}\xspace}         % {\Pa}{\Pb \Pc}

\def\to                 {\ensuremath{\rightarrow}\xspace}

\def\qsq       {{\ensuremath{q^2}}\xspace}

\def\Vcb  {{\ensuremath{V_{\cquark\bquark}}}\xspace}

\def\AT#1     {\ensuremath{A_{\mathrm{T}}^{#1}}\xspace}           % 2

\def\C#1      {\ensuremath{\mathcal{C}_{#1}}\xspace}                       % 9
\def\Cp#1     {\ensuremath{\mathcal{C}_{#1}^{'}}\xspace}                    % 7
\def\Ceff#1   {\ensuremath{\mathcal{C}_{#1}^{\mathrm{(eff)}}}\xspace}        % 9  
\def\Cpeff#1  {\ensuremath{\mathcal{C}_{#1}^{'\mathrm{(eff)}}}\xspace}       % 7
\def\Ope#1    {\ensuremath{\mathcal{O}_{#1}}\xspace}                       % 2
\def\Opep#1   {\ensuremath{\mathcal{O}_{#1}^{'}}\xspace}                    % 7

\newcommand{\tev}{\ifthenelse{\boolean{inbibliography}}{\ensuremath{~T\kern -0.05em eV}}{\ensuremath{\mathrm{\,Te\kern -0.1em V}}}\xspace}
\newcommand{\gev}{\ensuremath{\mathrm{\,Ge\kern -0.1em V}}\xspace}
\newcommand{\mev}{\ensuremath{\mathrm{\,Me\kern -0.1em V}}\xspace}
\newcommand{\kev}{\ensuremath{\mathrm{\,ke\kern -0.1em V}}\xspace}
\newcommand{\ev}{\ensuremath{\mathrm{\,e\kern -0.1em V}}\xspace}
\newcommand{\gevc}{\ensuremath{{\mathrm{\,Ge\kern -0.1em V\!/}c}}\xspace}
\newcommand{\mevc}{\ensuremath{{\mathrm{\,Me\kern -0.1em V\!/}c}}\xspace}
\newcommand{\gevcc}{\ensuremath{{\mathrm{\,Ge\kern -0.1em V\!/}c^2}}\xspace}
\newcommand{\gevgevcccc}{\ensuremath{{\mathrm{\,Ge\kern -0.1em V^2\!/}c^4}}\xspace}
\newcommand{\mevcc}{\ensuremath{{\mathrm{\,Me\kern -0.1em V\!/}c^2}}\xspace}

\def\mum  {\ensuremath{{\,\upmu\mathrm{m}}}\xspace}

\def\invfb   {\ensuremath{\mbox{\,fb}^{-1}}\xspace}

\newcommand{\chisq}{\ensuremath{\chi^2}\xspace}

\def\gsim{{~\raise.15em\hbox{$>$}\kern-.85em
          \lower.35em\hbox{$\sim$}~}\xspace}
\def\lsim{{~\raise.15em\hbox{$<$}\kern-.85em
          \lower.35em\hbox{$\sim$}~}\xspace}

\def\ptot       {\mbox{$p$}\xspace}
\def\pt         {\mbox{$p_{\mathrm{ T}}$}\xspace}

\def\evtgen     {\mbox{\textsc{EvtGen}}\xspace}

\def\geant      {\mbox{\textsc{Geant4}}\xspace}

\def\photos     {\mbox{\textsc{Photos}}\xspace}

\def\pythia     {\mbox{\textsc{Pythia}}\xspace}

\def\tell1  {TELL1\xspace}
\def\ukl1   {UKL1\xspace}

\usepackage{cite} % Allows for ranges in citations
\usepackage{mciteplus}

\def\csib{\xi _B}

\def\Sigmacpp  {{\ensuremath{\Sigmares^{++}_\cquark}}\xspace}
\def\Sigmacp  {{\ensuremath{\Sigmares^{+}_\cquark}}\xspace}
\def\Sigmacz    {{\ensuremath{\Sigmares^0_\cquark}}\xspace}
\def\rhosq{\rho ^2}
\def\Lcst{\ensuremath{\PLambda}_\cquark(2595)^+\xspace}%
\def\Lcsst{\ensuremath{\PLambda}_\cquark(2625)^+\xspace}
\def\Lchone{\ensuremath{\PLambda}_\cquark(2765)^+\xspace}
\def\Lchtwo{\ensuremath{\PLambda}_\cquark(2880)^+\xspace}
\def\Lcp{\ensuremath{\PLambda}_\cquark^+\xspace}
\def\Lcstar{\ensuremath{\PLambda}_\cquark^{\ast +}\xspace}
\def\mb{m_\bquark}
\def\mc{m_\cquark}
\def\pipi{\pip\pim}
\begin{document}

\renewcommand{\thefootnote}{\fnsymbol{footnote}}
\setcounter{footnote}{1}

% $Id: title-LHCb-PAPER.tex 78711 2015-08-06 07:54:32Z apuignav $
% ===============================================================================
% Purpose: LHCb-PAPER journal paper title page template
% Author: 
% Created on: 2010-09-25
% ===============================================================================

%%%%%%%%%%%%%%%%%%%%%%%%%
%%%%%  TITLE PAGE  %%%%%%
%%%%%%%%%%%%%%%%%%%%%%%%%
\begin{titlepage}
\pagenumbering{roman}

% Header ---------------------------------------------------
\vspace*{-1.5cm}
\centerline{\large EUROPEAN ORGANIZATION FOR NUCLEAR RESEARCH (CERN)}
\vspace*{1.5cm}
\noindent
\begin{tabular*}{\linewidth}{lc@{\extracolsep{\fill}}r@{\extracolsep{0pt}}}
\ifthenelse{\boolean{pdflatex}}% Logo format choice
{\vspace*{-2.7cm}\mbox{\!\!\!\includegraphics[width=.14\textwidth]{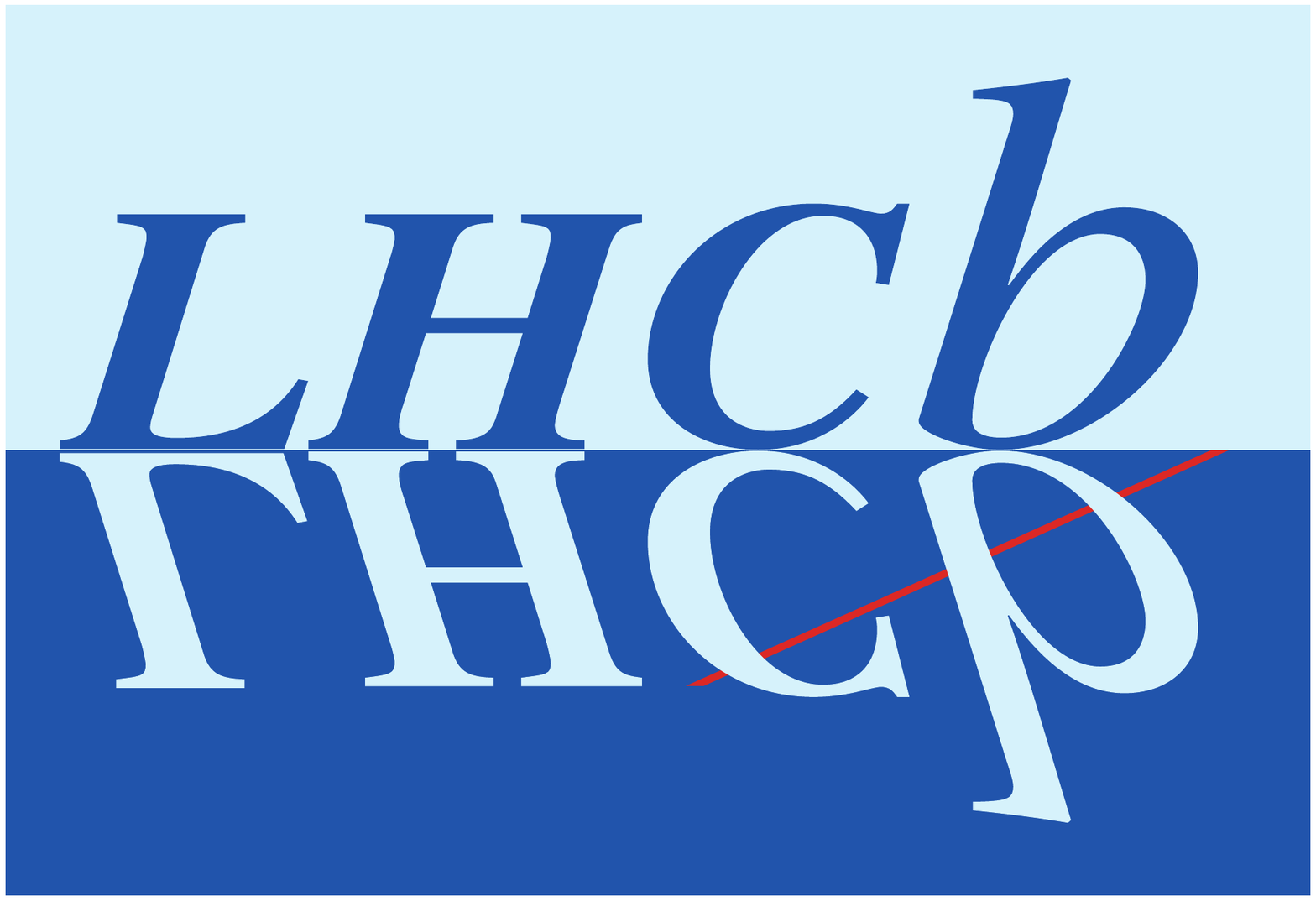}} & &}%
{\vspace*{-1.2cm}\mbox{\!\!\!\includegraphics[width=.12\textwidth]{lhcb-logo.eps}} & &}%
\\
 & & CERN-EP-2017-164 \\  % ID 
 & & LHCb-PAPER-2017-016 \\  % ID 
 & & \today \\ % Date - Can also hardwire e.g.: 23 March 2010
 & & \\
% not in paper \hline
\end{tabular*}

\vspace*{4.0cm}

% Title --------------------------------------------------
{\normalfont\bfseries\boldmath\huge
\begin{center}
 Measurement of the shape of the $\Lb\to\Lc \mun \neumb$ differential decay rate 
\end{center}
}

\vspace*{2.0cm}

% Authors -------------------------------------------------
\begin{center}
%In the footnote, replace 'paper' by 'letter' in case of submission to PRL or PLB 
The LHCb collaboration\footnote{Authors are listed at the end of this paper.}
\end{center}

\vspace{\fill}

% Abstract -----------------------------------------------
\begin{abstract}
  \noindent
  A measurement of the  shape of the differential decay rate and the associated Isgur-Wise function for the decay $\decay{\Lb}{\Lc\mun\neumb}$ is reported, using data corresponding to $3\invfb$  collected with the LHCb detector in proton-proton collisions. The  $\Lc\mun\neumb$(+ anything) final states are reconstructed through the detection  of a muon and a $\Lc$ baryon  decaying into $p\Km\pip$, and  the decays $\Lb\to\Lc\pip\pim\mun\neumb$ are used to determine contributions from   $\Lb\to \Lcstar\mun\neumb$ decays. The measured dependence of the differential decay rate upon  the squared four-momentum transfer between the heavy baryons, $\qsq$,  is compared with expectations from heavy-quark effective theory and from unquenched lattice QCD predictions. 
  
\end{abstract}

%\vspace*{2.0cm}

\begin{center}
  Published in  Phys. Rev. D
\end{center}

\vspace{\fill}

{\footnotesize 
\centerline{\copyright~CERN on behalf of the \lhcb collaboration, licence \href{http://creativecommons.org/licenses/by/4.0/}{CC-BY-4.0}.}}
\vspace*{2mm}

\end{titlepage}

%%%%%%%%%%%%%%%%%%%%%%%%%%%%%%%%
%%%%%  EOD OF TITLE PAGE  %%%%%%
%%%%%%%%%%%%%%%%%%%%%%%%%%%%%%%%

%  empty page follows the title page ----
\newpage
\setcounter{page}{2}
\mbox{~}
%\newpage
%
%% Author List ----------------------------
%%  You need to get a new author list!
%\input{LHCb_authorlist.tex}
%
%The author list for journal publications is provided by the Membership Committee shortly after 'approval to go to paper' has been given.
%%It will be made available on the page
%%\verb!http://www.physik.uzh.ch/~strauman/forMemCo/LHCb-PAPER-XXXX-XXX/! .
%It will be sent to you by email shortly after a paper number has beens assigned.
%The author list should be included already at first circulation, 
%to allow new members of the collaboration to verify whether they have been included correctly.
%Occasionally a misspelled name is corrected or associated institutions become full members.
%In that case, a new author list will be sent to you.
%In case line numbering doesn't work well after including the authorlist, try moving the \verb!\bigskip! after the last author to a separate line.
%
%
%The authorship for Conference Reports should be ``The LHCb
%  collaboration'', with a footnote giving the name(s) of the contact
%  author(s), but without the full list of collaboration names.

\cleardoublepage

\renewcommand{\thefootnote}{\arabic{footnote}}
\setcounter{footnote}{0}

\pagestyle{plain} 
\setcounter{page}{1}
\pagenumbering{arabic}

\section{Introduction}
In the Standard Model (SM) of particle physics, quarks participate in a rich pattern of flavor-changing transitions. The relevant couplings form a complex 3$\times$3  matrix, known as the Cabibbo-Kobayashi-Maskawa (CKM) matrix, characterized by just four independent parameters \cite{Rosner:1991nd}.  A vast body of measurements of individual CKM elements exists, and thus  the overall consistency of the SM picture of charged current interactions is highly over-constrained. Decades of experimental work have demonstrated the impressive consistency of experimental data with the CKM paradigm \cite{Charles:2015gya,Bona:2017cxr}; nonetheless,  the motivation to probe the CKM matrix remains strong. Effects of physics beyond the SM may be subtle, thus more precise measurements are necessary to unveil them. 
Semileptonic decays of heavy-flavored hadrons are commonly used to measure CKM parameters, as they involve only one hadronic current, parameterized in terms of scalar functions known as form factors.  The number of form factors needed to describe a particular decay depends upon the spin of the initial- and final-state hadrons \cite{Fakirov:1977ta,Bauer:1986bm}.  
A precise calculation of these form factors has been elusive for many years as it is not possible in perturbative QCD.   Heavy-Quark Effective Theory (HQET) provides the  framework to systematically include nonperturbative corrections in computations involving hadrons containing heavy quarks. In particular, in the limit of infinite heavy quark mass, all the form factors describing the semileptonic decay of a heavy-flavored hadron are proportional to a universal function, known as the Isgur-Wise (IW) function\cite{Isgur:1989vq}. Lattice QCD, namely the use of lattice formulations of QCD in large scale numerical
simulations,  has emerged in recent years as a technique with well defined and systematically improvable uncertainties which can
be applied to a wide range of processes and physical quantities\cite{Sachrajda:2016sck}. Predictions from the infinite heavy-quark mass limit are  useful as a check of  several Lattice  QCD calculations \cite{Mannel:2015osa}.

 The decay $\Lb\to\Lc\mun\neumb$ is described by six form factors corresponding to the vector and axial-vector components of the flavor-changing charged current \cite{Isgur:1990pm}. In HQET, $\Lb$ decays are particularly simple, as the light $\uquark\dquark$ quark pair has total spin $j=0$, and thus the chromomagnetic corrections, which are of the order of a few percent for $B$ mesons, are not present \cite{Bigi:2011gf}. In the static approximation of infinite heavy quark masses, the six form factors characterizing the baryonic semileptonic decay\footnote{The inclusion of charge-conjugate modes is implied throughout this paper.} $\Lb\to\Lc\mun\neumb$ can be expressed in terms of the elastic heavy-baryon Isgur-Wise function $\xi _B(w)$ \cite{Falk:1992wt}.  
The scalar invariant $w\equiv v_{\Lb}\cdot v_{\Lc}$  is related to the squared four-momentum transfer between the heavy baryons, $\qsq$, by 
\begin{equation}
w=(m_{\Lb}^{2}+m_{\Lc}^{2}-q^{2})/(2m_{\Lb}m_{\Lc}),
\label{eq:w}
\end{equation}       
where $v_{\Lb}$ and $v_{\Lc}$ are the four-velocities of the  $\Lb$ and $\Lc$ baryons, respectively, and $m_{\Lb}$ and $m_{\Lc}$ are the corresponding invariant masses. Nonperturbative corrections to the static limit can be expressed in terms of an expansion in powers of  1/$m_\cquark$ and 1$/m_\bquark$, where $m_\cquark$ and $m_\bquark$  represent the $c$- and $b$-quark masses respectively.  It has been shown in Ref. \cite{Georgi:1990ei} that the 1/$m_c$ term can be expressed in terms of $\xi _B(w)$ and  one dimensionful constant.  Moreover, partial cancellations  lead to small first-order corrections near $w=1$~\cite{Falk:1992ws}.

In the static approximation the differential  decay width of the $\Lb\to\Lc\mun\neumb$ decay is given by
\begin{equation}
\frac{d\Gamma}{dw}=GK(w)\xi^{2}_{B}(w),   
\label{eq:iw} 
\end{equation}where the constant factor $G$ is given by
\begin{equation}
G=\frac{2}{3}\frac{G_{F}^{2}}{(2\pi)^{3}}|V_{cb}|^{2}(m_{\Lb})^{4}r^{2}~~\text{with}~~r=m_{\Lc}/m_{\Lb} ,
\end{equation}
where $G_F$ represents the Fermi coupling constant \cite{Olive:2016xmw}, $|\Vcb|$ is the magnitude of the matrix element describing the coupling of the $\cquark$ quark to the $\bquark$ quark,  and the kinematic factor $K(w)$  is given by
\begin{equation}
K(w)=m_{\Lc}\sqrt{w^{2}-1}\ [3w(1-2rw+r^{2})+2r(w^{2}-1)].
\label{eq:kin}
\end{equation}
The function $\xi_B(w)$ cannot be determined from first principles in HQET, but calculations based on a variety of approaches exist.  The kinematic limit $w=1$ is special in HQET, as only modest corrections in the ($1/\mb$, $1/\mc)$ expansion are expected, due to the absence of hyperfine corrections \cite{Holdom:1993yn}. Thus it is interesting 
to express $\xi_B$ as a Taylor series expansion
\begin{equation}
\xi_B(w)=1-\rho^2 (w-1)+\frac{1}{2}\sigma^2 (w-1)^2+\dots  ,
\label{t-series}
\end{equation}
where $\rho^2$ is the magnitude of the slope of $\xi _B$ and $\sigma ^2$ is its curvature at $w=1$.
Sum rules provide constraints on $\rho^2$ and $\sigma^2$. In particular they require the slope at the zero recoil point  to be negative, and give bounds on the curvature and higher-order derivatives \cite{LeYaouanc:2003rn,Yaouanc:2009xe}. In addition they predict $\sigma^2\ge3/5[\rho^2+(\rho^2)^2]$ \cite{LeYaouanc:2008pq} and $\rho^2\ge3/4$. Table~\ref{tab:rhosq} summarizes  theoretical predictions for $\rho^2$ from quenched Lattice QCD, QCD sum rules, and a relativistic quark model.

Recently, state-of-the-art calculations of the six form factors describing the decay $\Lb\to\Lc\mun\neumb$ have been obtained using Lattice QCD with 2 + 1 flavors of dynamical domain-wall fermions  \cite{Detmold:2015aaa}. These form factors are calculated in terms of  $\qsq$. More details on this formalism are given in Appendix I. The resulting theoretical  uncertainty attached to a measurement of $|\Vcb|$ using this form factor prediction is  about 3.2\%. The precision of this calculation makes this approach an appealing alternative to the ones currently used, all based on $B$-meson semileptonic decays such as $\Bzb\to\Dstarp\mun\neumb$. Thus it is important to examine the model's agreement with measured quantities such as the shape of the $d\Gamma/d\qsq$ spectrum. 

The experimental knowledge of $\Lb$ semileptonic decays is quite sparse, as this baryon is too heavy to be produced at the $\ep\en\ B$-factories. The only previous experimental study of $\csib(w)$ was performed by the DELPHI experiment at LEP, which obtained $\rho^2= 2.03\pm 0.46\ {\rm(stat)} ^{+0.72}_{-1.00}\ {\rm(syst)}$, with an overall uncertainty of the order of  50\%\cite{Abdallah:2003gn}.

In this paper we describe a determination of the shape of the $w$ or $\qsq$ spectrum of the decay $\Lb\to\Lc\mun\neumb$ and compare it with functional forms related to a single form factor, inspired by  HQET, and the Lattice QCD prediction of Ref.~\cite{Detmold:2015aaa}.  
 Section~\ref{exp:method} presents the experimental procedure and simulated samples, while Sect.~\ref{sec:evreco} describes the method employed to reconstruct $\Lb\to\Lc\mun\neumb$ candidates and to estimate the corresponding kinematic variables $w$ and $\qsq$.  
 Section~\ref{raw:w} describes the method adopted to isolate the signal, the unfolding procedure used to account for experimental resolution effects, and the efficiency corrections. The fit results for $\xi_B(w)$ corresponding to different  functional forms  are summarized in Sect.~\ref{sec:dgdw}.  The same analysis procedure is used in Sect.~\ref{sec:dgdqsq} to derive the shape of the differential decay width  $d\Gamma/d\qsq(\Lb\to\Lc\mun\neumb)$ and  compare with the predictions of Ref.~\cite{Detmold:2015aaa}. These data are also fitted with a single form-factor parameterization that corresponds to  the HQET infinite heavy-quark mass limit.

\begin{table}
\begin{center}
\caption{Predictions for the  slope at zero recoil of the baryonic Isgur-Wise function $\csib$. The evaluation from Ref.~\cite{Ebert:2006rp} includes first-order corrections in HQET.}\label{tab:rhosq}
\begin{tabular}{lll}
\hline
 \multicolumn{1}{c}{$\rhosq$} & \multicolumn{1}{c}{Approach} & \multicolumn{1}{c}{Reference} \\
 \hline 1.35$\pm$0.13 & QCD sum rules & \cite{Huang:2005mea}\\
 1.2${^ {+0.8}_{ -1.1}}$ & Lattice QCD (static approximation) & \cite{Bowler:1997ej}\\[0.3ex]
1.51 & HQET + Relativistic wave function & \cite{Ebert:2006rp}\\ \hline
 & & \\
\end{tabular}
\end{center}
\vspace{-20pt}
\end{table}

\section{Experimental method}
\label{exp:method}
The data used in this analysis were collected with  the LHCb detector at the Large Hadron Collider at CERN and correspond to $1\invfb$ of integrated luminosity collected at a center-of-mass energy of $7\tev$ in 2011 and $2\invfb$ collected at a center-of-mass energy of $8\tev$ in 2012.

The LHCb detector~\cite{Alves:2008zz,LHCb-DP-2014-002}  is a single-arm forward spectrometer designed for the study of particles containing $\bquark$ or $\cquark$ quarks. The detector covers the  \mbox{pseudorapidity} range $2<\eta <5$, where $\eta$ is defined in terms of the polar angle $\theta$ with respect to the beam direction as 
$-\ln(\tan{\theta/2})$.  The detector includes a high-precision tracking system
consisting of a silicon-strip vertex detector surrounding the $pp$
interaction region~\cite{LHCb-DP-2014-001}, a large-area silicon-strip detector located upstream of a dipole magnet with a bending power of about
$4{\mathrm{\,Tm}}$, and three stations of silicon-strip detectors and straw
drift tubes~\cite{LHCb-DP-2013-003} placed downstream of the magnet.
The tracking system provides a measurement of the momentum, \ptot, of charged particles with a relative uncertainty that varies from 0.5\% at low momentum to 1.0\% at $200\gev$.\footnote{Natural units with $c$=$\hbar$=1 are used throughout.} The minimum distance of a track to a primary vertex, the impact parameter (IP), is measured with a resolution of $(15+29/\pt)\mum$,
where \pt is the component of the momentum transverse to the beam, in \gev.
Different types of charged hadrons are distinguished using information
from two ring-imaging Cherenkov detectors (RICH)~\cite{LHCb-DP-2012-003}. 
Photons, electrons and hadrons are identified by a calorimeter system consisting of
scintillating-pad and preshower detectors, an electromagnetic
calorimeter and a hadronic calorimeter. Muons are identified by a
system composed of alternating layers of iron and multiwire proportional chambers~\cite{LHCb-DP-2012-002}. The online event selection is performed by a trigger~\cite{LHCb-DP-2012-004}, which consists of a hardware stage, based on information from the calorimeter and muon systems, followed by a software stage, which applies a full event reconstruction. 

 Muon candidates are first required to pass the hardware trigger that selects muons with a transverse momentum $\pt>1.6$ ($1.8$)$\gev$ for the 2011 (2012) data taking period. In the subsequent software trigger, events with one particle identified as a muon are selected if at least
  one of the final-state particles has both
  $\pt>0.8\gev$ and IP larger than $100\mum$ with respect to all
  of the primary $pp$ interaction vertices~(PVs) in the
  event. In the offline selection, trigger signals are associated with reconstructed particles. Selection requirements can therefore be made on the trigger selection itself and on whether the decision was due to the signal candidate, other particles produced in the pp collision, or a combination of both. This classification of trigger selections can be used for data-driven efficiency determination. Finally, the tracks of two or more of the final-state
  particles are required to form a vertex that is significantly
  displaced from the PVs.

 Our study makes use of simulated semileptonic decays, where $pp$ collisions are generated using 
\pythia~\cite{Sjostrand:2006za,*Sjostrand:2007gs} 
 with a specific \lhcb
configuration~\cite{LHCb-PROC-2010-056}.  Decays of hadronic particles
are described by \evtgen~\cite{Lange:2001uf}, in which final-state
radiation is generated using \photos~\cite{Golonka:2005pn}. The
interaction of the generated particles with the detector, and its response,
are implemented using the \geant
toolkit~\cite{Allison:2006ve, *Agostinelli:2002hh} as described in
Ref.~\cite{LHCb-PROC-2011-006}.

\section{Event reconstruction}
\label{sec:evreco}
To isolate a sample of $\Lb\to \Lc\mun\neumb X$  semileptonic decays, where $X$ represents the undetected  particles produced with the $\Lc$ in the $c$-quark hadronization, we combine $\Lc\to p\Km\pip$ candidates with tracks identified as muons. We consider candidates where a well-identified muon passing the hardware and software trigger algorithms with momentum greater than $3\gev$  is found. 
Charmed baryon candidates are formed from hadrons with momenta greater than $2\gev$ and transverse momenta greater than $0.3\gev$. In addition we require that the average of the magnitudes of the transverse momenta of the hadrons forming the $\Lc$ candidate be greater than $0.7\gev$. Kaons, pions, and protons are identified using the RICH system. Each track's IP significance with respect to the associated primary vertex is required to be greater than 9.\footnote{The associated primary vertex to a $\Lb\to \Lc\mun\neumb X$ candidate is selected as the primary vertex which minimizes the IP significance of the $\Lc\mun$ system.} Moreover, the selected tracks must be consistent with coming from a common vertex:  the $\chisq$ per  degree of freedom ($\chisq$/DOF) of the vertex fit must be smaller than 6. In order to ensure that the direction of the parent $\Lb$ is well measured, the $\Lc$ vertex must be distinct from the primary $pp$ interaction vertex. To this end, we require that the flight-distance significance of the $\Lc$ candidate (defined as the measured flight distance divided by its uncertainty) with respect to the associated PV be greater than 100.

\begin{figure}[b]
\begin{center}

\includegraphics[width=3 in]{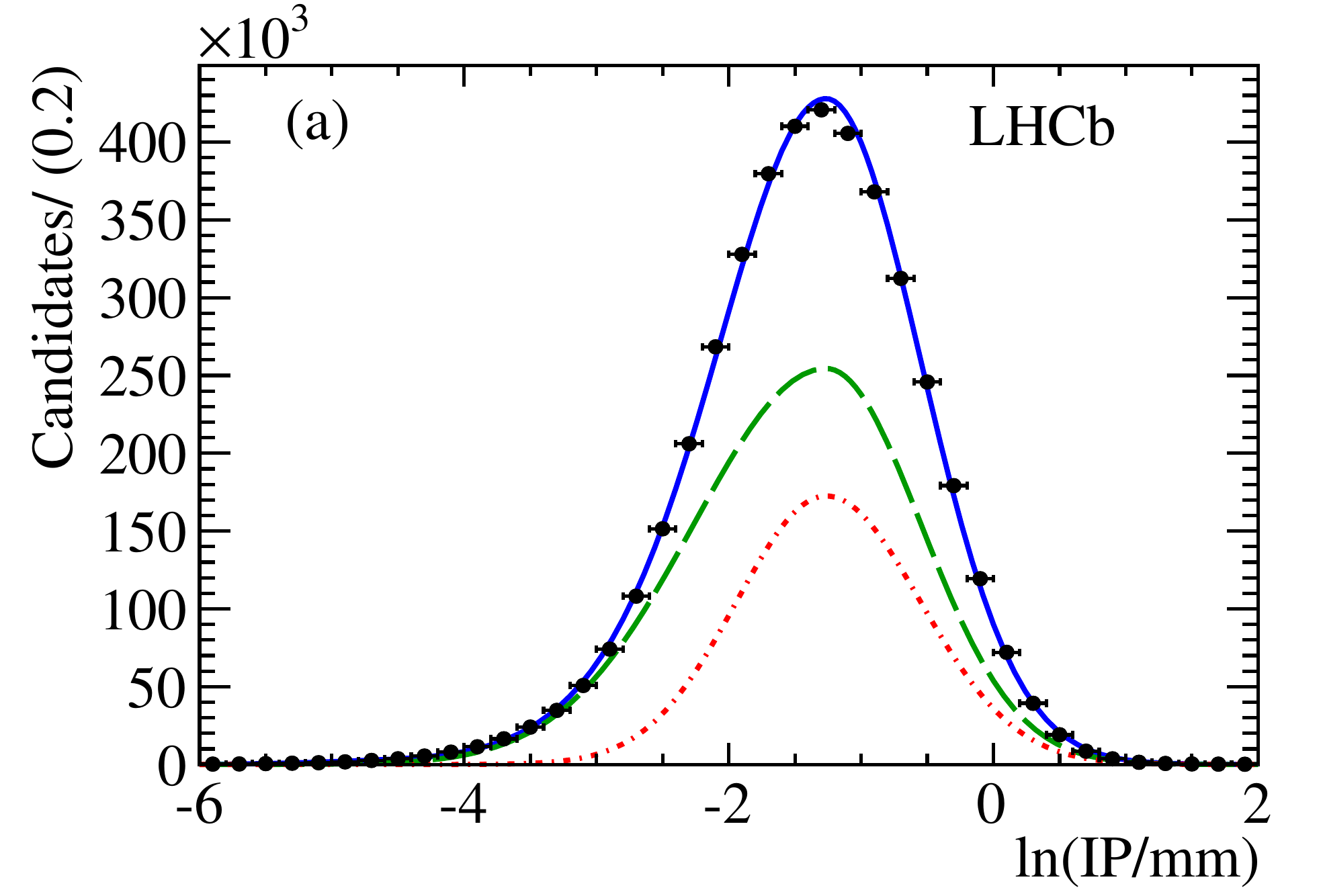}\includegraphics[width=3 in]{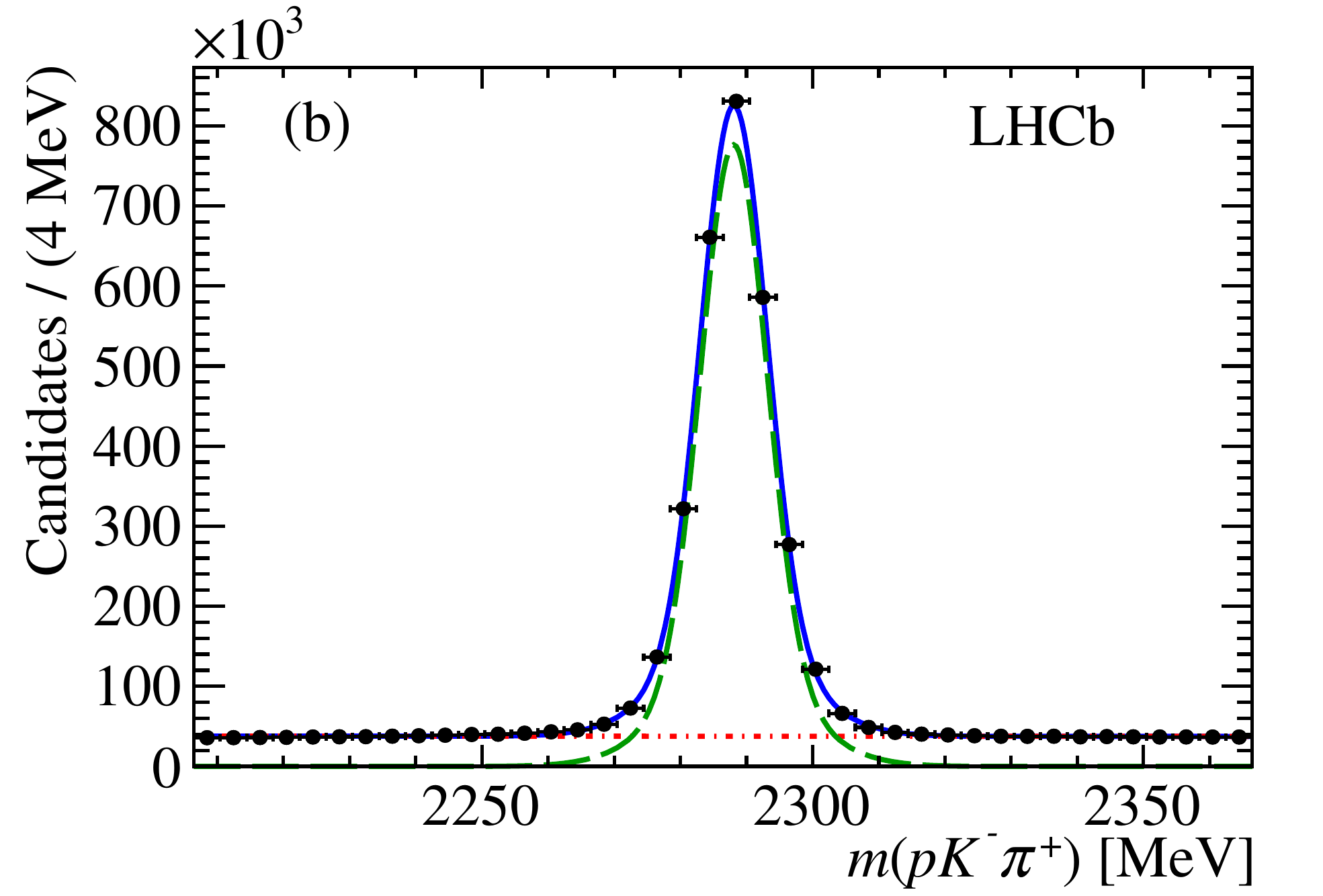}
\end{center}
\caption{(a) The  ln(IP/mm)  distribution and (b) $p\Km\pip$ invariant mass for $\Lc$ candidate combinations with a
muon. The red (dashed-dotted) curves show the combinatorial $\Lc$ background, the green (dashed) curves the $\Lc$ from $\Lb$ and the blue-solid curves the total yields.} \label{Lcall}
% \end{center}
\end{figure}

 Partially reconstructed $\Lb$ baryon candidates are formed combining $\mun$  and  $\Lc$ candidates  which are consistent with coming from a common vertex, and  we require that the angle between the direction of the momentum of the $\Lc\mun$ candidate and the line from the associated PV to the $\Lc\mun$ vertex  be less than 45 mrad. As the $\Lc$ baryon is a $\Lb$ decay product with a small but significant lifetime, we require that the difference in the component of the decay vertex position of the charmed hadron candidate along the beam axis and that of the beauty candidate  be positive. We explicitly require that the $\Lb$ hadron candidate pseudorapidity be between 2 and 5. We measure $\eta$ using the line defined by connecting the associated PV and the vertex formed by the $\Lc$ and the $\mun$ lepton. Finally, the invariant mass of the $\Lc\mun$ system must be between $3.3$ and $5.3\gev$.
These selection criteria ensure that the $\Lc$ candidates  are decay products of $\Lb$ semileptonic decays. In particular the background from directly produced $\Lc$ (prompt $\Lc$) is highly suppressed. This is quantified by an unbinned extended maximum likelihood fit to the two-dimensional $pK^-\pi^+ $ invariant mass and ln(IP/mm) distributions of the $\Lc$ candidates, where  ``/mm" refers to the length unit used to measure the IP. The ln(IP/mm) component allows us to determine the small prompt $\Lc$ background. The parameters of the IP distribution of the prompt sample are found by examining directly-produced charm hadrons, as described in Ref.~\cite{Aaij:2010gn}. An empirical probability density function (PDF) derived from  simulation is used for the $\Lc$ from $\Lb$ component. We find $(2.74\pm 0.02)\times 10^6$ $\Lc\to p\Km\pip$ candidates, which can be interpreted as $\Lb\to\Lc\mun\neumb X$ decays, and  we determine the prompt $\Lc\to p\Km\pip$ fraction to be  1.5\%, which can be neglected. The corresponding fit  is shown in Fig.~\ref{Lcall}.

%\afterpage{\clearpage}
 
Our goal is the study of the ground-state semileptonic decay $\Lb\to\Lc\mun\neumb$, thus we need to estimate the contributions from $\Lcstar$ decaying into $\Lc\pi\pi$ states. Theoretical predictions suggest that the inclusive rate $\Lb\to\Lc\mun\neumb X$ is dominated by the exclusive channel $\Lb\to\Lc\mun\neumb$ \cite{Albertus:2004wj,Pervin:2005ve}.  The residual contribution is expected to be  accounted for by the $\Lb\to\Lcst\mun\neumb$ and $\Lb\to\Lcsst\mun\neumb$ channels. Other modes, such as $\Lb\to\Sigmacp\mun\neumb$,   are  suppressed in the static limit and to order $1/m_Q$,  where $m_Q$ represents the heavy quark mass ($m_\cquark$ or $m_\bquark$) \cite{Leibovich:1997az}, with an additional stronger suppression factor of the order $ (m_\dquark- m_\uquark)/m_\cquark$ rather than  $(m_\dquark- m_\uquark)/m_{\Lambda_{\rm QCD}}$ \cite{Isgur:1990pm}. 

We use $\Lb\to\Lc\pip\pim\mun\neumb$ decays to infer contributions from the excited $\Lc$ modes, where the $\Lc$ candidates are selected as $p\Km\pip$  combinations whose invariant mass is within $\pm 20\mev$ of the nominal $\Lc$ mass. The $\Lc\mun\neumb$ candidates  are combined with  pairs of opposite-charge pions that satisfy criteria similar to those used to select the pions from the $\Lc$ decay. The minimum transverse momentum of these pions is  required to be $0.2\gev$ and the transverse momentum of the $\Lc\pip\pim$ system is required to be greater than $1.5\gev$. Lastly, the $\chisq$ per degree of freedom of the vertex fit for the $\Lc\pip\pim$ system must be smaller than 6.

\begin{figure}[b]
\begin{center}
\includegraphics[width=3 in]{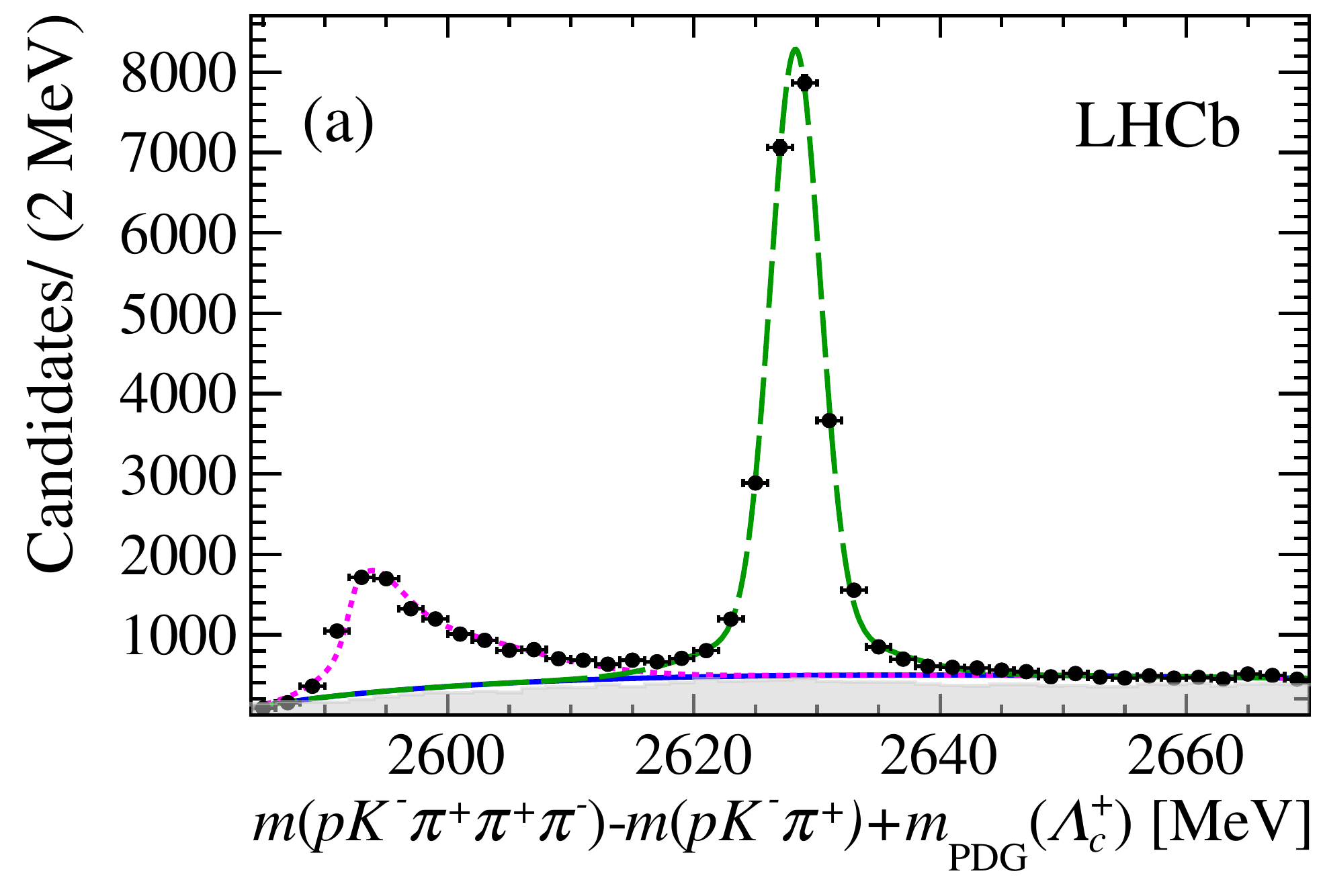}\includegraphics[width=3 in]{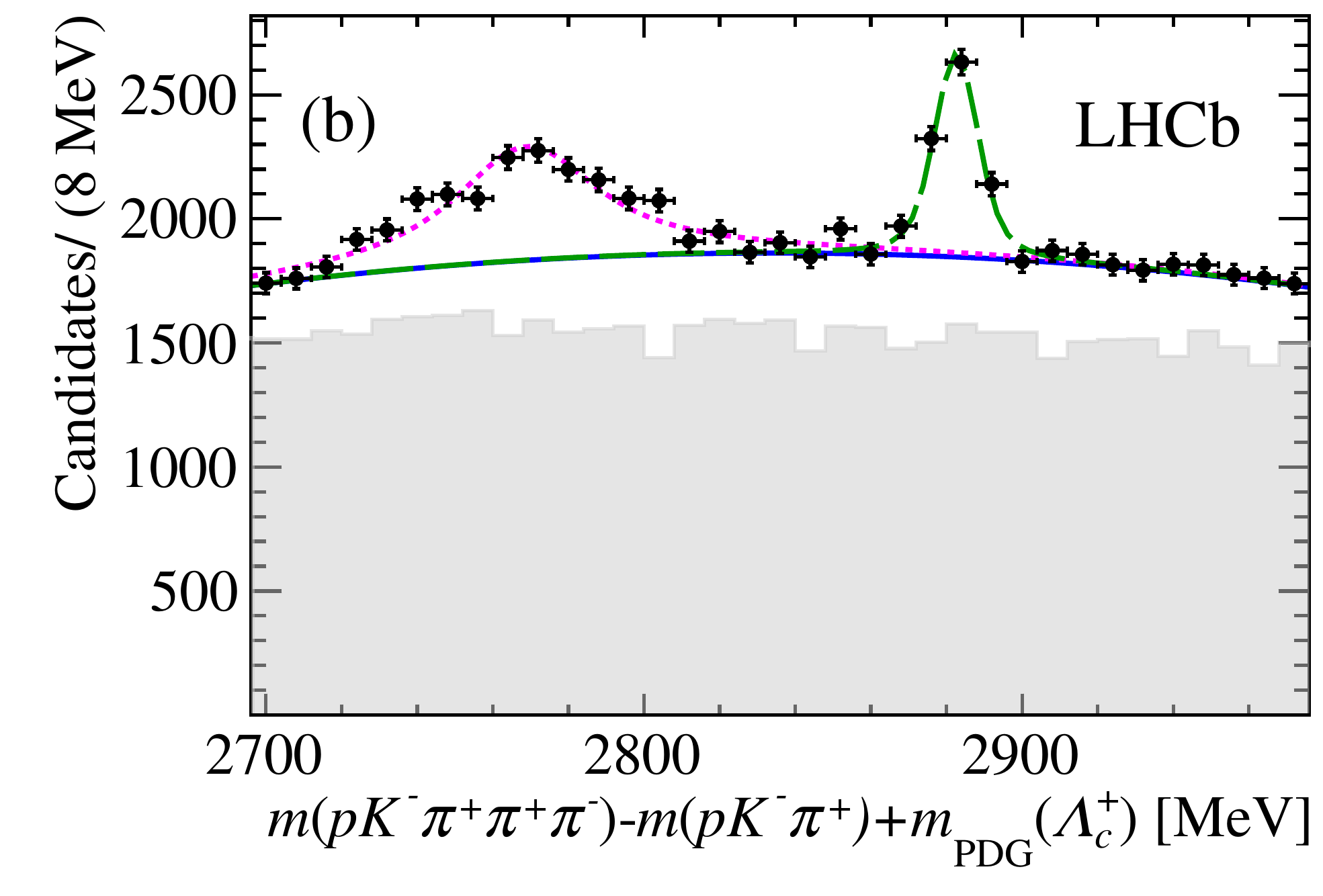}
\end{center}
\caption{The mass difference $m(p\Km\pip\pip\pim)-m(p\Km\pip)$ added to the known $\Lc$ mass, $m_{\rm PDG}(\Lc)$ ~\cite{Olive:2016xmw}, for candidates with $p\Km\pip$ invariant mass within $\pm$20 MeV of the known $\Lc$ mass in candidate semileptonic decays for the entire $w$ range:  data are shown as black dots, the combinatoric background is shown as a blue solid line, and the gray histogram shows the WS spectrum, obtained by combining a   $\pip\pip$ or $\pim\pim$ pair  with $\Lc$ instead of $\pipi$.  The signal fits are identified as follows: (a) for $m<2700$ MeV, the $\Lcst $ as magenta dashed line,  and the $\Lcsst$  as green long-dashed line, (b) for $m>2700$ MeV,  the $\Lchone $ as magenta dashed line,  and the $\Lchtwo$  as green long-dashed line.}  \label{fig:lcstar-mass}
\end{figure}

The resulting spectrum, measured as the mass difference $m(p\Km\pip\pim\pip)-m(p\Km\pip)$ added to the known $\Lc$ mass\cite{Olive:2016xmw},  is shown in Fig.~\ref{fig:lcstar-mass}. We see peaks corresponding to the $\Lcst $, $\Lcsst$, $\Lchone$, and $\Lchtwo$ resonances. The  $\Lcst$ is only a few MeV above the kinematic threshold  and thus it is not well described by a Breit-Wigner function. The baseline fit for this resonance uses a PDF consisting of the sum of two bifurcated Gaussian functions. As a check, we  use an S-wave relativistic Breit-Wigner   convolved with a Gaussian function with standard deviation $\sigma =2$ MeV that accounts for the detector resolution. While the second parameterization is more accurate, the fits to the invariant mass spectra in different kinematic bins are more stable with the baseline parameterization. 
We fit the $\Lcsst$ signal with a double Gaussian PDF with shared mean, as the natural width is expected to be well below the measured detector resolution. The shape of the combinatoric background  PDF is inferred from wrong-sign (WS) candidates, where  a $\pip\pip$ or $\pim\pim$ pair is combined with $\Lcp$ instead of $\pipi$. In addition, we observe peaks corresponding to two higher mass resonances, with masses and widths consistent with the $\Lchone$ and $\Lchtwo$ baryons\cite{Olive:2016xmw}. In order to determine their yields, we fit the two signal peaks with single Gaussian PDFs with unconstrained masses and widths. The measured yields for the four $\Lc$ final states, as well as the $\Lc\mun\neumb X$ final state, are presented in Table~\ref{tab:pipiyields}.  
%\afterpage{\clearpage}

\begin{table}
\begin{center}
\caption{Measured raw yields for the four $\Lcstar\mun\neumb$ final states and the inclusive $\Lc\mun\neumb X$. }\label{tab:pipiyields}
{\small
\begin{tabular}{cc}
\hline
 \multicolumn{1}{c}{Final state}  & \multicolumn{1}{c}{Yield} \\
 \hline 
$\Lcst\mun\neumb $ & \hphantom{0}8569 $\pm$ 144\\
$\Lcsst\mun\neumb $ & 22965 $\pm$ 266\\
$\Lchone\mun\neumb $ & \hphantom{0}2975 $\pm$ 225\\
$\Lchtwo\mun\neumb $ & 1602 $\pm$ 95\\
\hline
$\Lc\mun\neumb X$ & \hphantom{0}(2.74$\pm$0.02)$\times 10^6$\\
\hline
%$\Lc\mun\neumb X$ d $(2.74\pm 0.02)\times 10^6$ & 
\end{tabular}}
\end{center}
\end{table}
\newpage

 The measured contributions from the two heavier $\Lcstar$ final states, shown in Table~\ref{tab:pipiyields}, are smaller than those from $\Lb\to\Lcst\mun\neumb $ and $\Lb\to\Lcsst\mun\neumb $ decays. No theoretical prediction for nonresonant  $\Lb\to\Lc\pip\pim\mun\neumb X$ exists, but we  estimate  systematic uncertainties due to the subtraction of  this component with an alternative fit of the $\Lb\to\Lc\pip\pim\mun\neumb X$ spectrum from candidate $\Lb$ semileptonic decays, where we derive both the yield and shape of the combinatoric background from the WS sample.  

The kinematical quantities $\qsq$ and $w$  in the decay $\Lb\to\Lc\mun\neumb$ can be calculated if the magnitude of the $\Lb$ momentum is known.  The $\Lb$ flight direction can be inferred from the primary and secondary vertex locations, and this input, combined with the constraints from energy and momentum conservation,  implies the following relationship for  $p_{\Lb}$

\begin{eqnarray}
\label{eq:pb}
\left[\left(\frac
{\hat{p}_{\Lb}\cdot \vec{p}_{\Lc\mun}}{E_{\Lc\mun}}
\right)^2 -1\right] p_{\Lb}^2+
\left[(m_{\Lb}^2+m_{\Lc\mun}^2)\frac{\hat{p}_{\Lb}\cdot\vec{p}_{\Lc\mun}}{E^2_{\Lc\mun}}\right]p_{\Lb}\\\nonumber
+\left[\left(\frac{(m_{\Lb}^2+m_{\Lc\mun}^2)}{2E_{\Lc\mun}}\right)^2-m_{\Lb}^2\right]=0,
\end{eqnarray}
where the unit vector $\hat{p}_{\Lb}$ is the direction of the $\Lb$ baryon, $\vec{p}_{\Lc\mun}$ is the momentum of the $\Lc\mun$ pair, $E_{\Lc\mun}$ is the energy of the $\Lc\mun$ pair, $m_{\Lc\mun}$ is the invariant mass of the $\Lc\mun$ pair,  $m_{\Lb}$ is the nominal mass of the $\Lb$ baryon, and $\Lc$  identifies the  $p\Km\pip$ combination.
This is a quadratic equation, reflecting the lack of knowledge of the neutrino orientation in the $\Lb$ rest frame with respect to the $\Lb$ direction in the laboratory. The solution corresponding to the lower value of $p_{\Lb}$, which is correct between 50\% and 60\% of the time depending upon the kinematics of the final state, is chosen in the $\qsq$ and $w$ determination as simulation studies have shown that this choice introduces the smallest bias.   The $w$ resolution is estimated from simulated data in different $w$ intervals.  The distributions of  differences between reconstructed and generated $w$  are fitted with double-Gaussian functions and the effective standard deviations are found to be between 0.01 and 0.05.  The overall $w$ resolution is estimated with a fit with a triple-Gaussian function, and has an effective standard deviation $\sigma$ equal to $0.028$.

\section{\hspace{-10pt}The spectral distribution  \boldmath${dN_{\rm corr}/dw(\Lb\to\Lc\mun\neumb)}$}
\label{raw:w}
 The $\Lb\to\Lc\mun\neumb X$ candidates are separated into 14 equal-size bins of reconstructed $w$ in the full kinematic range $1\le w \le 1.43$.  The parameters of the  PDFs describing the signal and background components are determined from the fit to the overall $p K^-\pip$ mass spectrum. The contributions from semileptonic decays including higher-mass baryons in the final state is evaluated by fitting the $\Lc\pip\pim$ mass spectra with two different methods. In the first, we fit for the four resonances shown in Fig.~\ref{fig:lcstar-mass} using a PDF derived from the WS sample to model the background, and then use the simulation to correct for efficiency. In the second, we determine the signal yields of the $\Lcstar$ states by subtracting the  WS background and treating the residual smooth component of the spectrum as originating from a semileptonic decay $\Lb\to\Lc\mun\neumb X$. The second method provides an  estimate of the systematic uncertainty introduced by  the contribution from nonresonant $\Lc\pip\pim $ components of the hadron spectrum, as the smooth component of this spectrum is likely to comprise also combinatoric background.
  
 Next, we correct the raw $\Lc\mun\neumb X$ and $\Lc\pip\pim\mun\neumb X$ signal yields for the corresponding software trigger efficiencies, which are derived with a data-driven method \cite{LHCb-DP-2012-004}, based on the determination of $\Lc\mun\neumb X$ events where a positive trigger decision is provided by the signal candidates, and events where the trigger decision is independent of the signal. Then we subtract the raw yields  reported in Table~\ref{tab:pipiyields}, scaled by the corresponding efficiency ratios $\varepsilon (\Lb\to\Lc\mun\neumb X)\over \varepsilon (\Lb\to\Lc\pip\pim\mun\neumb X)$, from the corrected $\Lc\mun\neumb X$ yields. These ratios are derived from $\Lb\to\Lcst\mun\neumb$  and  $\Lb\to\Lcsst\mun\neumb$ simulations. The higher mass yields are scaled by an average of these two corrections, as no model for these semileptonic decays is available.  These corrections account for the efficiency loss due to the reconstruction of the additional pion pairs, as well as for the unseen $\Lb\to\Lc\piz\piz\mun\neumb X$ decay and are only mildly dependent upon the invariant mass of the final state. The expectation is that  $\Lb\to\Lc\pip\pim\mun\neumb$ accounts for  two-thirds of the  inclusive dipion final state. We have checked this prediction by studying the inclusive final states $\Sigmacpp \mun \neumb X$, $\Sigmacp \mun\neumb X$, and $\Sigmacz\mun\neumb X$. Taking into account the difference in the $\Lcp\pip\pim\mun\neumb X$ and $\Lcp\piz\mun\neumb X$ detection efficiencies, estimated with simulations, we measure the ratio $R=N(\Lcp\pip\pim)/N(\Lcp\pip\pim+\Lcp\piz\piz)$ with
\begin{equation}
R=\frac{N(\Sigmacpp)+N(\Sigmacz)}{N(\Sigmacpp)+N(\Sigmacz)+N(\Sigmacp)\cdot [\varepsilon(\Lcp\pip\pim\mu)/\varepsilon(\Lcp\piz\mu)]} ,
\label{eq:R}
\end{equation}
where $N(\Sigmacpp)$ and $N(\Sigmacz)$ are the detected yields for the final states $\Sigmacpp\pim\mu\nu$ and $\Sigmacz\pip\mu\nu$, $N(\Sigmacp)$ is the detected yield for the final state $\Sigmacp\mu\nu X$ and $\varepsilon(\Lcp\pip\pim\mu)/\varepsilon(\Lcp\piz\mu)$ is the ratio between the reconstruction efficiencies of these final states calculated with  simulation. A simulation study gives $\varepsilon(\Lcp\pip\pim\mu)/\varepsilon(\Lcp\piz\mu)=25.9\pm  2.7$, where the uncertainty reflects the limited sample size of the simulation. We obtain $R= 0.63\pm 0.14$, where the statistical uncertainty is due to limited $\piz$ reconstruction efficiency, consistent with the expectation $R=2/3$, and a negligible $\Sigmacp\mun\neumb$ component in the denominator of Eq.~\ref{eq:R}.

\begin{figure}[t]
\begin{center}

\includegraphics[width=3 in]{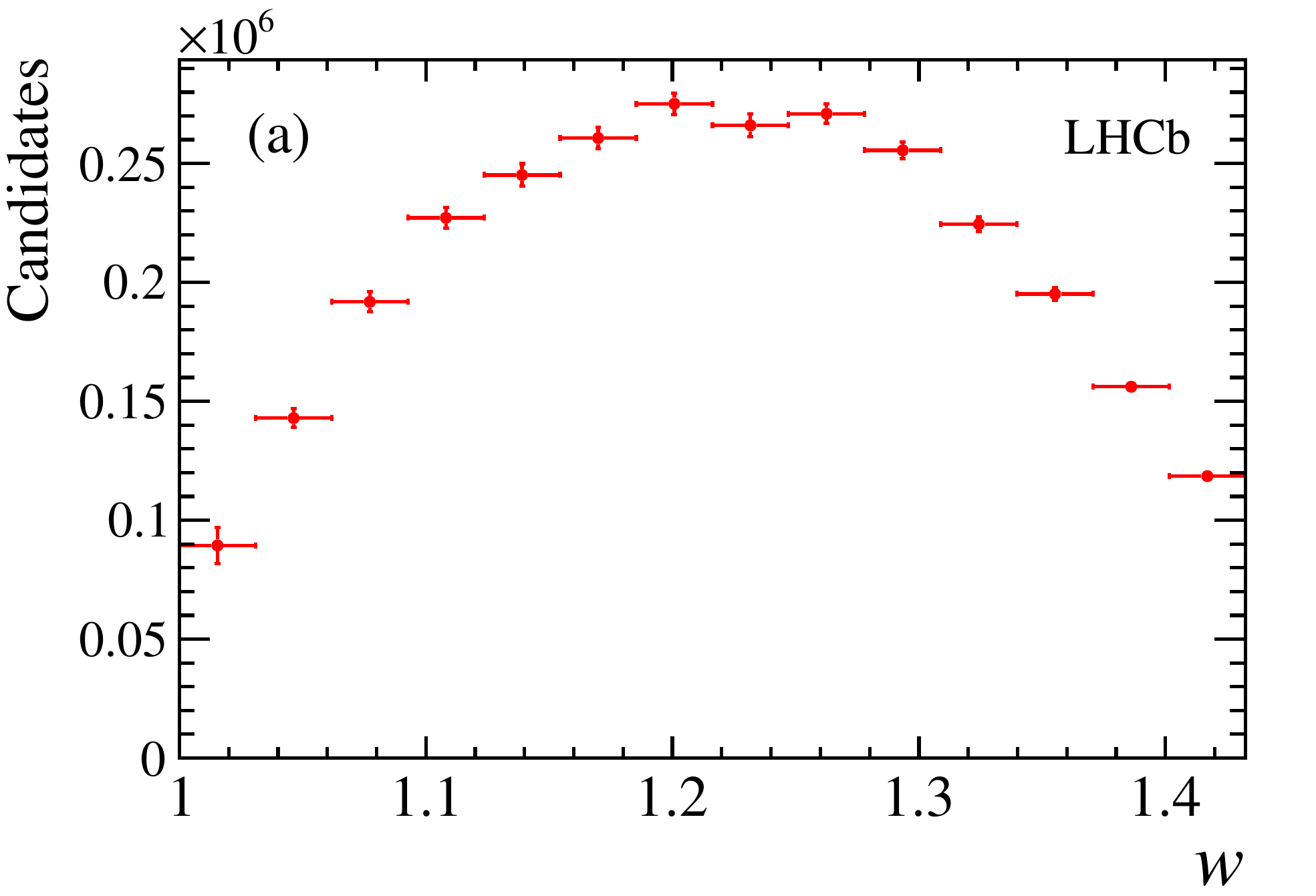}\includegraphics[width=3 in]{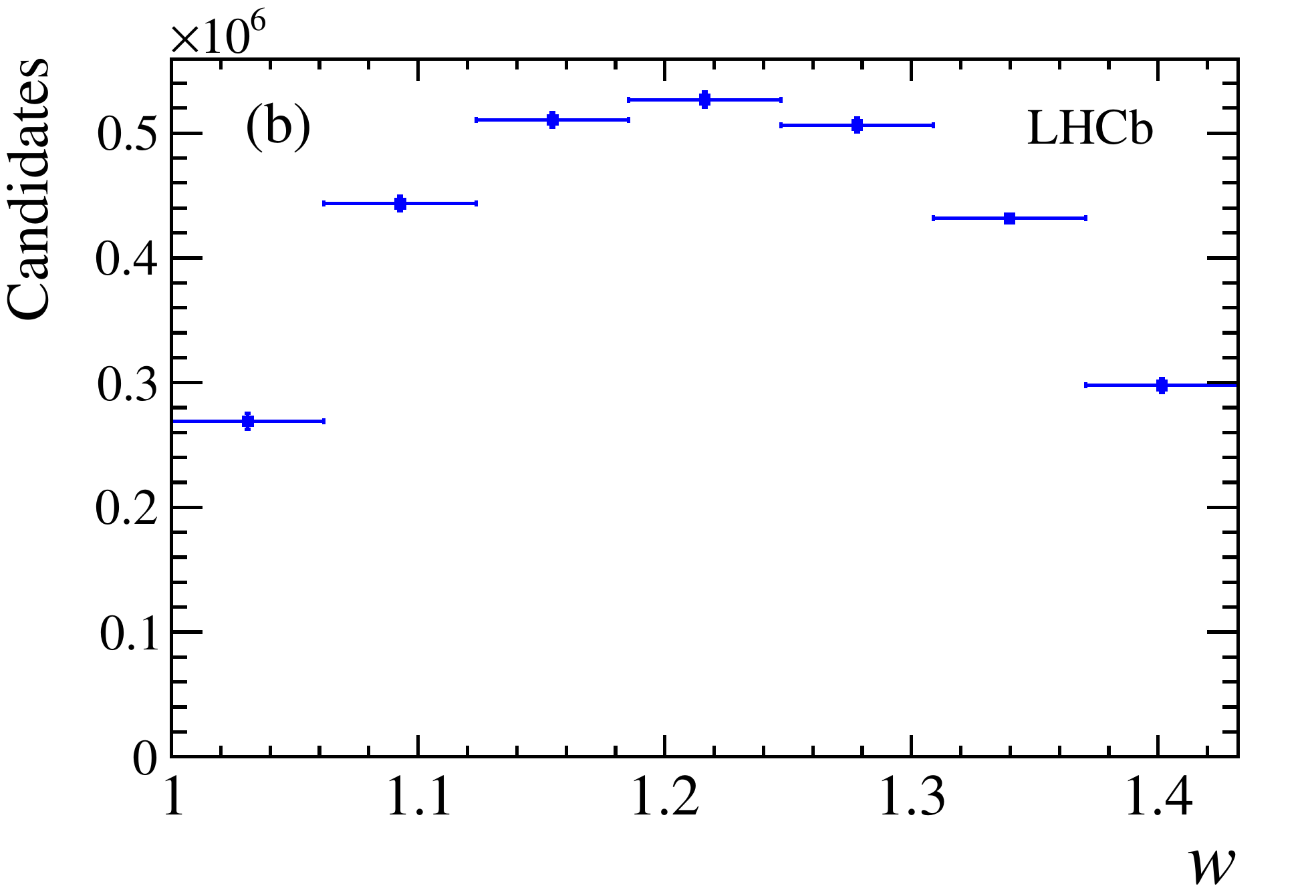}
\end{center}
\vspace*{-8mm}
\caption{ The spectra (a) $dN_{\rm meas}/dw$ before unfolding  and  (b) $dN_{\rm u}/dw$ after unfolding, for the decay $\Lb\to\Lc\mun\neumb$. The latter spectrum is then corrected for acceptance and reconstruction efficiency and fitted to the IW function $\xi_B(w)$ with the procedure discussed in the text.}
\label{fig:unfold}
\end{figure}

The $\Lb\to\Lc\mun\neumb$ spectrum $dN_{\rm meas}/dw$ is then unfolded to account for the detector resolution and other $w$ smearing effects such as the possible choice of the wrong solution of Eq.~\ref{eq:pb}. The procedure adopted is based on  the single value decomposition (SVD) method \cite{Hocker:1995kb} using the RooUnfold package\cite{Adye:2011gm}.  We  choose to divide the unfolded spectrum $dN_{\rm u}/dw$ into seven $w$ bins, to be consistent with the recommendation of Ref. \cite{blobel:school} to divide the measured spectrum into a number of bins at least twice as many as the ones in the corrected spectrum. The SVD method includes a regularization procedure that depends upon a parameter $k$\cite{Hocker:1995kb}, ranging between unity and the number of degrees of freedom, in our case 14. Simulation studies demonstrate that $k=4$ is optimal in our case. Variations associated with different choices  of $k$ have been studied and are included in the systematic uncertainties.  We have performed closure tests with different simulation models of the $\Lb\to\Lc\mun\neumb$ dynamics, and verified that this unfolding procedure does not bias the reconstructed distribution. The spectra before and after unfolding are shown in Fig.~\ref{fig:unfold}. Finally,  using simulated samples of signal events, we correct the unfolded spectrum for $w$-dependent acceptance and selection efficiency to obtain the distribution $dN_{\rm corr}/dw$. Various kinematic distributions have been studied in simulation and data and we find that they are all in good agreement.  

%\newpage

%\input{ff-iw}
\section{The shape of \boldmath$\xi_B(w)$ for $\Lb\to \Lc\mun\neumb$ decays}
\label{sec:dgdw}
 In order to determine the shape of the Isgur-Wise  function $\xi_B(w)$, we  use  the square root of $dN_{\rm corr}/dw$ divided by the kinematic factor $K$($\langle w \rangle$), defined in Eq.~\ref{eq:kin},  evaluated at the midpoint in the seven unfolded $w$ bins.  We derive the IW shape with a $\chisq$ fit, where the $\chisq$ is formed using the full covariance matrix of  $dN_{\rm corr}/dw$. 

\begin{figure}[t]
\begin{center}
\includegraphics[height=2in]{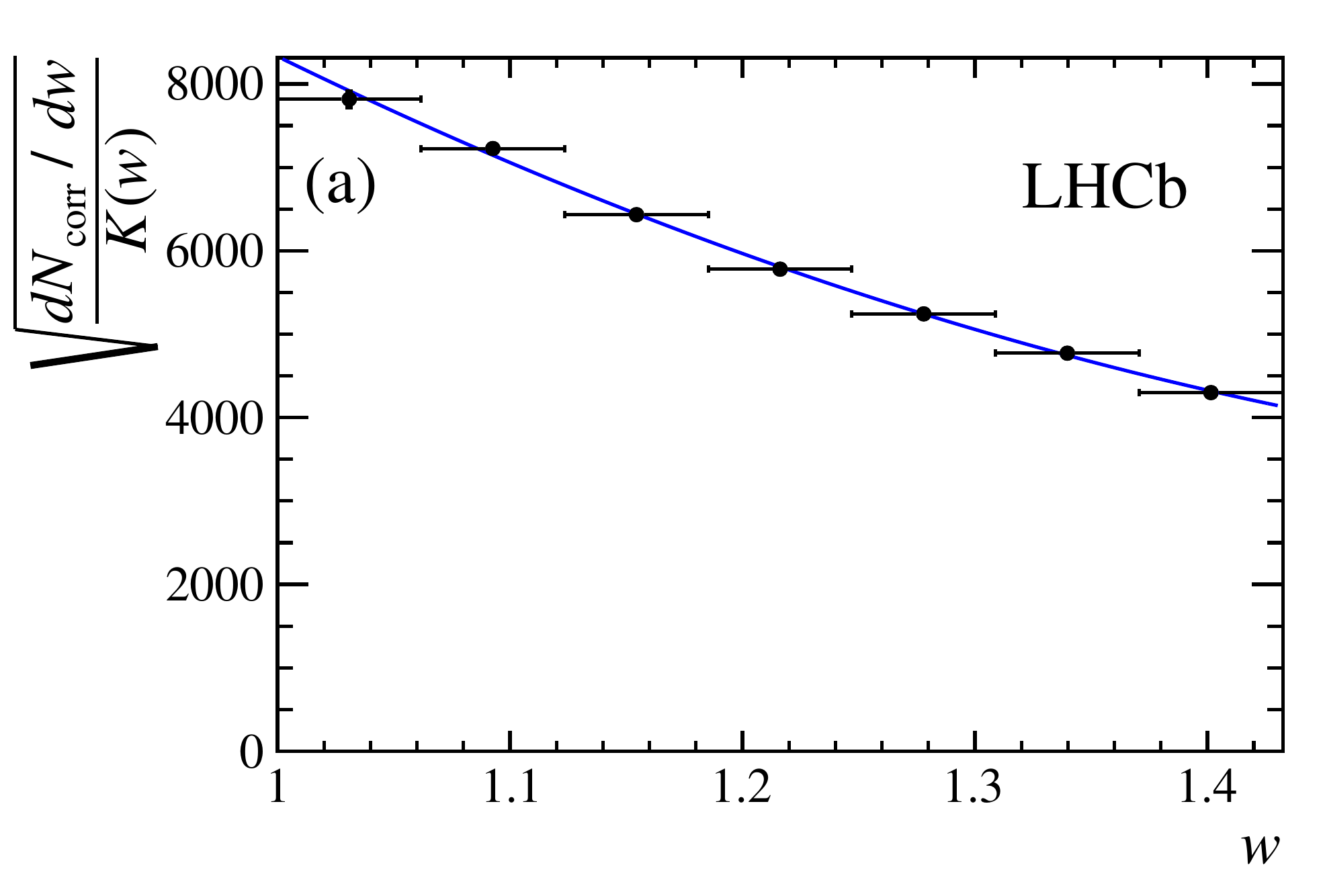} \includegraphics[height=2in]{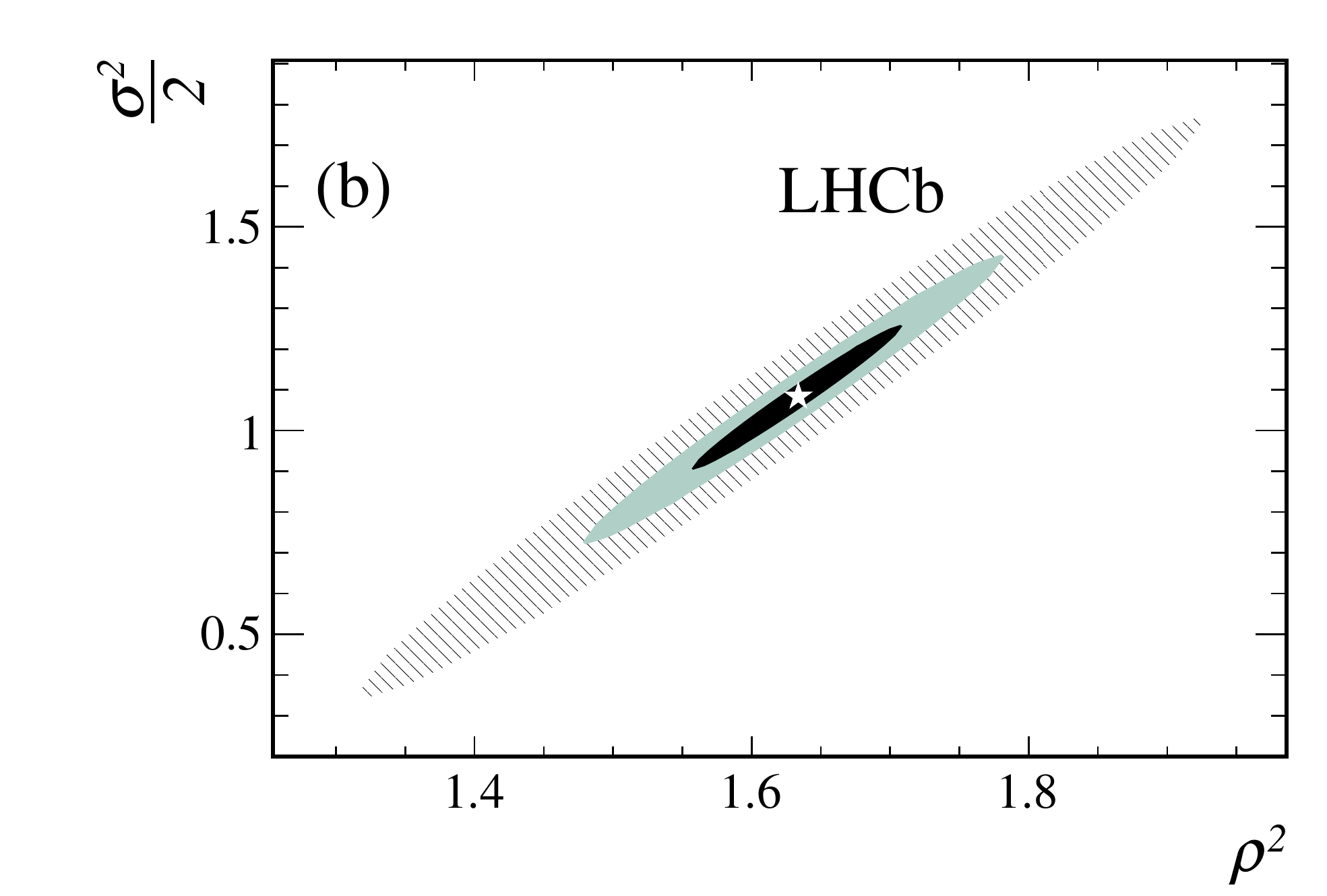} 
\end{center}
\caption{(a) The Isgur-Wise function fit for the decay $\Lb\to\Lc\mun\nu$ with a Taylor series expansion in $(w-1)$ up to second order.  The black dots show the data and the solid (blue) line shows the fitted function with the second-order Taylor series expansion model. The vertical scale is in arbitrary units.  (b)  The correlation between slope $\rho^2$ and curvature $\sigma ^2 /2$: the three ellipses correspond to the 1$\sigma$, 2$\sigma$, and 4$\sigma$ contours. }
\label{fig:taylor}
\end{figure}

We use various functional forms to extract the slope, $\rhosq$, and curvature, $\sigma^2$, of $\xi_B(w)$. The first functional form  is motivated by the 1/$N_c$ expansion \cite{Jenkins:1992se},  where $N_c$ represents the number of colors, and has an exponential shape parameterized as
\begin{equation}
\xi_{B}(w)=\exp[-\rho^{2}(w-1)].
\end{equation}
The second functional form, the so called ``dipole" IW function,  which is more consistent with sum-rule bounds \cite{Yaouanc:2009xe}, is given by
\begin{equation}
\xi_{B}(w)=\left( \frac{2}{w+1} \right)^{2\rho^2}.
\end{equation}
Finally, we can use a simple Taylor series expansion of the Isgur-Wise function and fit for the slope and curvature parameters using the Taylor series expansion introduced in Eq.~\ref{t-series}. Figure~\ref{fig:taylor} shows the measured $\xi_B(w)$ and the fit results with this parameterization.
Table~\ref{tab:shape-par} summarizes the slope and curvature at zero recoil obtained with the three fit models. Note that the curvature is an independent parameter only in the last fit, while in the first two models it is related to the second derivative of the IW function. 

\newpage

\begin{table}[b]
  \caption{Summary of the values for the slope and curvature of the Isgur-Wise function with different parameterizations. The quoted uncertainties are statistical only. The models marked with ``*" have only the slope at zero recoil as a free parameter, thus the curvature is derived from the fitted $\rho^2$.}
  \begin{center}\begin{tabular}{lcccc}
  \hline
  \multicolumn{1}{c}{Shape} & $\rho ^2$ & $\sigma ^2$ & correlation coefficient&$\chisq$/ DOF\\
  \hline
  Exponential* & 1.65 $\pm$ 0.03& 2.72 $\pm$ 0.10& 100\% &5.3/5\\
  Dipole* & 1.82 $\pm$ 0.03 & 4.22 $\pm$ 0.12 & 100\% & 5.3/5\\
  Taylor series & 1.63 $\pm$ 0.07&2.16 $\pm$ 0.34 &97\% & 4.5/4 \\
  \hline
   \end{tabular}\end{center}
\label{tab:shape-par}
\end{table}

As the slope of the IW function is the most relevant quantity to determine $|\Vcb|$ in the framework of HQET \cite{Falk:1992ws}, we focus our studies on the systematic uncertainties on this parameter. We consider several sources of systematic uncertainties, which are listed in Table~\ref{tab:systematics}. The first two are determined by changing the fit models for $\Lc$ and $\Lcst$ and $\Lcsst$ signal shapes in the corresponding candidate mass spectra. The software trigger efficiency uncertainty is estimated by using an alternative procedure to evaluate this efficiency using the trigger emulation in the LHCb simulation. In order to assess systematics associated with the bin size, we perform the analysis with different binning choices. The sensitivity to the $\Lb\to\Lc\mun\neumb$ form factor modeling is assessed by reweighting the simulated $w$ spectra to correspond to  different $\xi_B$ functions with slopes ranging from 1.5 to 1.7. The ``phase space averaging" sensitivity is estimated by comparing the fit to the expression for $dN_{\rm corr}/dw$   with the fit to $1/K(\langle w \rangle)\sqrt{dN_{\rm corr}/dw}$. The uncertainty associated with the $\Lb\to\Lcstar \mun\neumb$  modeling is evaluated by changing the relative fraction of  $\Lc\pip\pim$ versus $\Lc\piz\piz$ of the $\Lcstar$ spectrum by $\pm$ 20\%.  Finally, we use the alternative evaluation of the fraction of $\Lb\to\Lc\pip\pim\mun\neumb$ which includes the maximum possible nonresonant  component to assess the sensitivity to residual $\Lcstar$ components in the subtracted spectrum. The total systematic uncertainty in $\rhosq$ is 0.08.

The value of $\rhosq$ obtained from the Taylor series expansion is $$\rhosq=1.63\pm 0.07 \pm 0.08,$$  which is consistent with Lattice calculations~\cite{Bowler:1997ej}, QCD sum rules~\cite{Huang:2005mea}, and relativistic quark model~\cite{Ebert:2006rp} expectations. The measured slope is compatible with theoretical predictions and with the bound
$\rho^2\ge$ 3/4\cite{LeYaouanc:2003rn}. The measured curvature  $\sigma ^2$ is compatible within uncertainties with the lower bound 
$\sigma^2\ge3/5[\rho^2+(\rho^2)^2]$ \cite{LeYaouanc:2008pq}. 

\begin{table}[b]
	\caption{Summary of the systematic uncertainties on the slope parameter $\rho^2$. The total uncertainty is obtained by adding the individual components in quadrature.}
	\begin{center}\begin{tabular}{l|c}
			\hline
			Source & $\sigma(\rho^2)$ \\
			\hline
			Signal fit for $\Lc$ & $0.02$\\
			Signal PDF for $\Lcstar$& $0.02$ \\
			Software trigger efficiency &  $0.02$ \\ 
			$w$ binning & $0.03$\\
			SVD unfolding regularization & $0.03$\\
			Phase space averaging & 0.03 \\
			$ \Lb\to\Lc\mun\neumb$  modeling & $0.03$ \\
			$\Lb\to\Lcstar \mun\neumb$ modeling & $0.03$ \\
			Additional components of the semileptonic spectrum & $0.02$\\ 
			$\Lb$ kinematic dependencies & 0.02 \\
			\hline
			Total & $0.08$ \\
			\hline
		\end{tabular}\end{center}
		\label{tab:systematics}
	\end{table}

\section{Comparison with unquenched lattice predictions}
\label{sec:dgdqsq}

The lattice QCD calculation in Ref.~\cite{Detmold:2015aaa} uses a helicity-based description of the six form factors governing  $\Lb\to\Lambdares$   transitions introduced in Ref.~\cite{Feldmann:2011xf}. The calculation uses state-of-the-art techniques encompassing the entire $\qsq$ region. The stated uncertainties on the predicted width are therefore larger than what is expected in a high-$\qsq$ region, but remain rather small, namely 6.3\%. This corresponds to a 3.2\% theoretical uncertainty on $|\Vcb|$, thus raising the prospect  of an additional precise independent determination of $|\Vcb|$. 

The simplest check on this theoretical prediction consists of a comparison of the predicted shape of $d\Gamma/d\qsq$ and the measured data. Thus we measure the distribution  $dN_{\rm corr}/d\qsq$  with the same procedure adopted to derive $dN_{\rm corr}/dw$, including efficiency corrections and the unfolding procedure, with the same number of bins used to determine the raw and unfolded yields. We produce seven corrected yields and their associated covariance matrix, where the nondiagonal terms are related to the unfolding procedure. 
We then perform a $\chisq$ fit to the seven experimental $dN_{\rm corr}/d\qsq$ data points using the theoretical functional shape given in Eq. 85 of Ref.~\cite{Detmold:2015aaa}, which also provides the nominal values of the form factor parameters, and thus we leave only the relative normalization floating. This fit uses a covariance matrix that combines experimental and theoretical uncertainties, which yields a $\chisq$ equal to 1.32 for 6 degrees of freedom, and a corresponding p-value of  97\%. This shows that the predicted shape is in good agreement with our measurement.

\begin{figure}[hbt]
\begin{center}
\includegraphics[width=3.4  in]{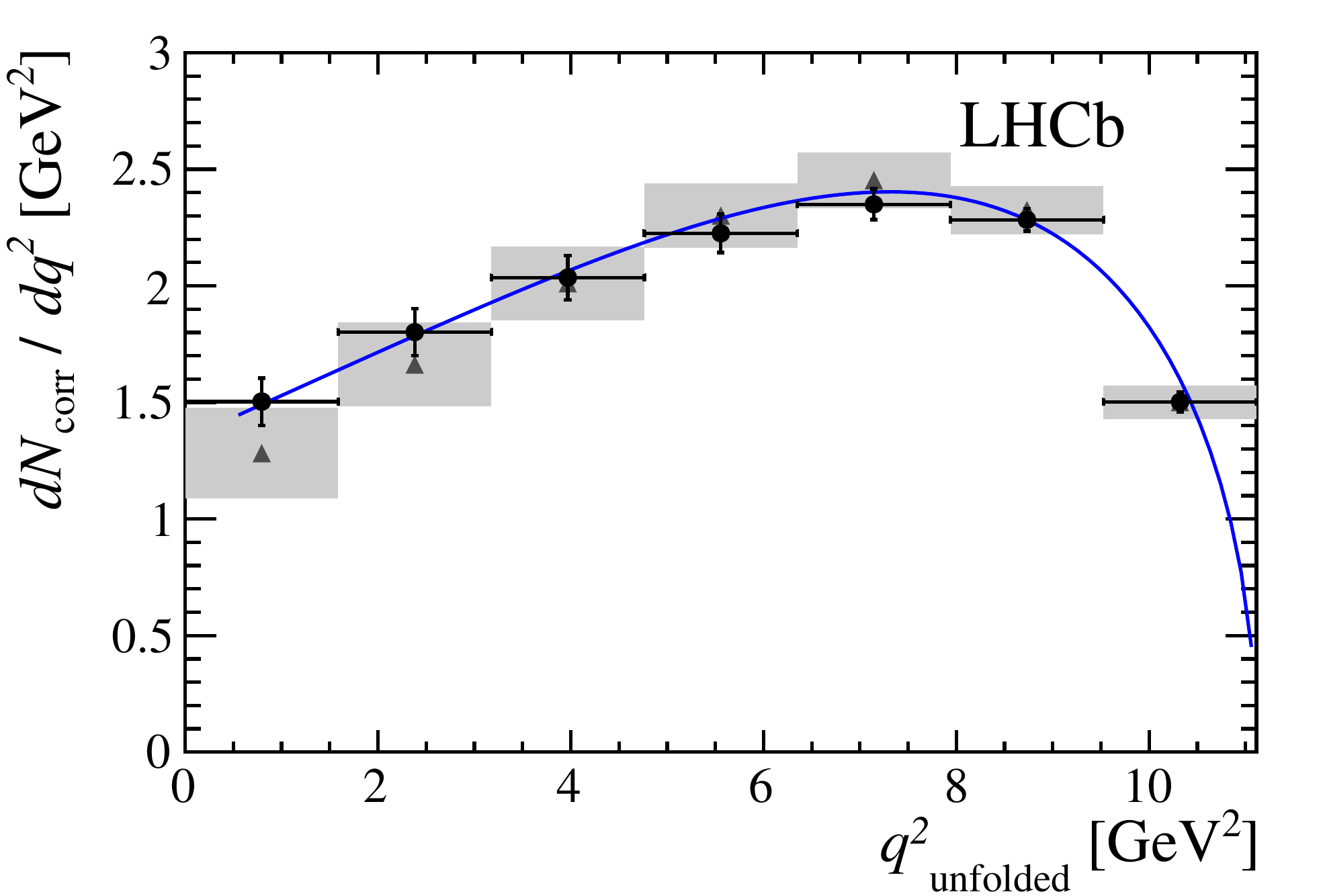}  
\end{center}
\caption{Comparison between the fit to the seven experimental data points using either the Lattice QCD calculation of Ref.~\cite{Detmold:2015aaa}, shown as grey points with a shaded area corresponding to the binned 1$\sigma$ theory uncertainty, or a single form factor fit in the $z$-expansion, shown as the solid blue curve. The data points, modulo a scale factor, are shown as black points with error bars. }
\label{fig:qsqfits}
\end{figure}

The form factor decomposition in Ref.~\cite{Detmold:2015aaa} does not allow a straightforward extrapolation to the HQET limit of infinite heavy-quark masses. However, we know that in the static limit all the form factors are proportional to a single universal function. In order to assess how well our data are consistent with the static limit, we perform a second $\chisq$ fit assuming that all the form factors are proportional to a single $z$-expansion function~\cite{Hill:2006ub}. Fits with different pole masses used in the six form factors determined in Ref.~\cite{Detmold:2015aaa} are performed. The overall shape does not change appreciably;  the pole mass of $6.768\gev$ is preferred.  The two fit parameters are the coefficients $a_0$ and $a_1$, giving the strength of the first two terms in the $z$-expansion. The resulting fitted shape is shown in Fig.~\ref{fig:qsqfits}. This fit has a $\chisq$ equal to 1.85 for 5 degrees of freedom, with a corresponding p-value of 87\%. Note that the shape obtained with a single form-factor is very similar to the one predicted in Ref.~\cite{Detmold:2015aaa}. This is consistent with the HQET prediction \cite{Holdom:1993yn} that the shape of the differential distribution is well described by the static approximation, modulo a scale correction of the order of 10\%, reflecting higher-order contributions. Further details of this fit and the fit using the Lattice QCD calculation can be found in the Appendix.

%\newpage

%\input{conclusions}

\section{Conclusions}
A precise measurement of the shape of the Isgur-Wise function describing the semileptonic decay $\decay{\Lb}{\Lc \mun \neumb}$ has been performed.   The measured slope is consistent with theoretical models and the bound
$\rho^2\ge 3/4$ \cite{LeYaouanc:2003rn}. The measured curvature  $\sigma ^2$ is consistent with the lower-bound constraint  $\sigma^2\ge3/5[\rho^2+(\rho^2)^2]$ \cite{LeYaouanc:2008pq}.
The shape of $d\Gamma/d\qsq$ is  studied and found to be well described by the unquenched lattice QCD prediction of Ref.~\cite{Detmold:2015aaa}, as well as by a single form-factor parameterization.  Further studies with a suitable normalization channel will lead to a precise independent determination of the CKM parameter $|\Vcb|$.

\section*{Acknowledgements}

\vspace{1cm}

\noindent We express our gratitude to our colleagues in the CERN
accelerator departments for the excellent performance of the LHC. We
thank the technical and administrative staff at the LHCb
institutes. We acknowledge support from CERN and from the national
agencies: CAPES, CNPq, FAPERJ and FINEP (Brazil); MOST and NSFC (China);
CNRS/IN2P3 (France); BMBF, DFG and MPG (Germany); INFN (Italy); 
NWO (The Netherlands); MNiSW and NCN (Poland); MEN/IFA (Romania); 
MinES and FASO (Russia); MinECo (Spain); SNSF and SER (Switzerland); 
NASU (Ukraine); STFC (United Kingdom); NSF (USA).
We acknowledge the computing resources that are provided by CERN, IN2P3 (France), KIT and DESY (Germany), INFN (Italy), SURF (The Netherlands), PIC (Spain), GridPP (United Kingdom), RRCKI and Yandex LLC (Russia), CSCS (Switzerland), IFIN-HH (Romania), CBPF (Brazil), PL-GRID (Poland) and OSC (USA). We are indebted to the communities behind the multiple open 
source software packages on which we depend.
Individual groups or members have received support from AvH Foundation (Germany),
EPLANET, Marie Sk\l{}odowska-Curie Actions and ERC (European Union), 
Conseil G\'{e}n\'{e}ral de Haute-Savoie, Labex ENIGMASS and OCEVU, 
R\'{e}gion Auvergne (France), RFBR and Yandex LLC (Russia), GVA, XuntaGal and GENCAT (Spain), Herchel Smith Fund, The Royal Society, Royal Commission for the Exhibition of 1851 and the Leverhulme Trust (United Kingdom).

%\input{lattice-supplementary}
%\titleformat{\section}{\large\bfseries}{\appendixname~\thesection .}{0.5em}{}
%\begin{appendix}
%\appendix
\section*{Appendix I: Analytical expression for \boldmath$d\Gamma/d\qsq$}
\label{sec:appendix}

This Appendix describes the formalism used in the $d\Gamma/d\qsq$ fits. In particular, we give the expression of $d\Gamma/d\qsq$ in terms of the form factor basis chosen in Ref.~\cite{Detmold:2015aaa},  the so-called ``helicity form factors".  In addition, we  show the corresponding expression used to model the static limit.

\begin{table}[b]
	\caption{Masses of the relevant form factor poles in the physical limit (in GeV).}
	\begin{center}
		\begin{tabular}{ccccccc}
			\hline
			$f$     & \hspace{1ex} & $J^P$ & 
			%\hspace{1ex}  & \hspace{1ex} 
			& $m_{\rm pole}^f(\Lb \to \Lambda_c)~ [\gev]$ & \hspace{1ex} & \\
			\hline
			$f_+$, $f_\perp$ && $1^-$ && 6.332  \\
			$f_0$            && $0^+$ && 6.725 \\
			$g_+$, $g_\perp$ && $1^+$ && 6.768 \\
			$g_0$            && $0^-$ && 6.276  \\
			\hline\end{tabular}
	\end{center}
	\label{tab:polemasses}
\end{table}

The Lattice QCD calculations reported  in Ref.~\cite{Detmold:2015aaa} predict the differential decay width  $d\Gamma(\Lb\to \Lc\mun\neumb)/d\qsq$  as follows
\begin{eqnarray}
\nonumber \frac{\mathrm{d}\Gamma}{\mathrm{d} q^2} &=&
\frac{G_F^2 |\Vcb| ^2 \sqrt{s_+s_-}
}{768 \pi ^3 m_{\Lb}^3 } \left(1-\frac{m_\ell^2}{q^2}\right)^2 \\
\nonumber     &&\times\Bigg\{4 \left(m_\ell^2+2 q^2\right) \left(
s_+ \left[  g_\perp  \right(\qsq)]^2 + s_- \left[f_\perp(\qsq)\right]^2 \right)  \\
\nonumber    && \hspace{2ex} +2 \frac{m_\ell^2+2 q^2}{q^2} \left(s_+
\left[\left(m_{\Lb}-m_X\right) g_+(\qsq) \right]^2+s_- \left[\left(m_{\Lb}+m_X\right)f_+(\qsq) \right]^2\right)\\
&&\hspace{2ex} +\frac{6 m_\ell^2}{q^2} \left(s_+ \left[ \left(m_{\Lb}-m_X\right) f_0(\qsq)
\right]^2 + s_-
\left[ \left(m_{\Lb}+m_X\right)g_0(\qsq)  \right]^2\right)\Bigg\},
\label{eq:dgdq2}
\end{eqnarray}
where $g_\perp$, $f_\perp$, $g_+$, $f_+$, $g_0$, and $f_0$ represent the six form factors necessary to describe this decay, $X\equiv\Lambda_c$ denotes the final-state baryon, $m_\ell$ represents the mass of the muon, $\qsq$ is the squared  four-momentum transfer between the heavy baryons, and
\begin{equation}
s_\pm =(m_{\Lb} \pm m_X)^2-q^2.
\end{equation}
The six form factors are cast in terms of the $z$-expansion~\cite{Hill:2006ub} up to first order, and  have the functional form
\begin{equation}
f(\qsq)= \frac{1}{1-\qsq/(m^{f}_{\rm pole})^{2}}\times\left[ a_0^f+a_1^fz^{f}(\qsq)\right],
\label{eq:firstorder}
\end{equation}
where $z^f(\qsq)$ is given by
\begin{eqnarray}
z^f(q^2) &=& \frac{\sqrt{t^{f}_+-q^2}-\sqrt{t_+^{f}-t_0}}{\sqrt{t_+^{f}-q^2}+\sqrt{t_+^{f}-t_0}}, \\
t_0 &=& (m_{\Lb} - m_{X})^2,
\end{eqnarray}
and $t_+^{f }$ is given by
\begin{equation}
t_+^{f }= (m^{f}_{\rm pole})^2, 
\end{equation}
and the pole masses used in the calculations are shown in Table~\ref{tab:polemasses}. The parameters $a_0^f$ and $a_1^f$ for the six form factors describing this decay are given in Table VIII of Ref.~\cite{Detmold:2015aaa}.

In the static limit all the helicity form factors are proportional to a single universal function. Thus, we use a common $z$-expansion parameterization 
\begin{eqnarray}
	\nonumber \frac{\mathrm{d}\Gamma}{\mathrm{d} q^2} &=&
	\frac{G_F^2  |\Vcb |^2 \sqrt{s_+s_-}
	}{768 \pi ^3 m_{\Lb}^3 } \left(1-\frac{m_\ell^2}{q^2}\right)^2g_{\perp}^{2}(\qsq) \\
	\nonumber     &&\times\Bigg\{4 \left(m_\ell^2+2 q^2\right) \left(s_+ + s_-\right)  \\
	   &&\hspace{2ex} +\frac{4}{q^2}\left[s_+\left(m_{\Lb}-m_X\right)^2 + s_-\left(m_{\Lb}+m_X\right)^2\right] \left[2m_\ell^2+q^2\right]\Bigg\},
	\label{eq:singlez}
\end{eqnarray}
where the choice of $g_{\perp}$ reflects the choice of the pole mass used in the single $z$-expansion fit given in Sect.~\ref{sec:dgdqsq}. We have performed the fits with various choices of pole masses and examined the effects on the shape $d\Gamma/d\qsq$ and found the shape did not vary significantly, though it was found that the parameters defining $g_{\perp}$ yielded the optimal fit. In this case, the fit parameters are the coefficients $a_0$ and $a_1$ in the $z$-expansion parameterization of $g_{\perp}(\qsq)$, which has the form shown in Eq.~\ref{eq:firstorder}.

\section*{Appendix II: Measured normalized spectra $dN_{\rm corr}/d\qsq$ and associated covariance matrix}

In this appendix we report the seven measured data points $dN_{\rm corr}/d\qsq$ and the corresponding covariance matrix, shown in Table~\ref{tab:ndata} and Table~\ref{tab:covmat} respectively.
\begin{table}[h]
\begin{center}
\caption{Measured normalized yields  $dN_{\rm corr}(\Lb \to \Lc\mun\neumb)/d\qsq$ . }\label{tab:ndata}
{\small
\begin{tabular}{c|c}
\hline
$\qsq$ [GeV$^2$] & $dN_{\rm corr}/d\qsq$\\
 \hline 
0.80 & 1.50 $\pm$ 0.10 \\
2.38 & 1.80 $\pm$ 0.10 \\
3.97 & 2.04 $\pm$ 0.10 \\
5.56 & 2.23 $\pm$ 0.08 \\
7.14 & 2.35 $\pm$ 0.07 \\
8.73 & 2.28 $\pm$ 0.05 \\
10.32 & 1.50 $\pm$ 0.04 \\ 
\hline
%$\Lc\mun\neumb X$ d $(2.74\pm 0.02)\times 10^6$ & 
\end{tabular}}
\end{center}
\end{table}

\begin{table}[h]
\begin{center}
\caption{Covariance matrix of the measured normalized yields  ${\rm Cov}[dN_{\rm corr}(\Lb \to \Lc\mun\neumb)/d\qsq$] . }\label{tab:covmat}
{\small
\begin{tabular}{c|ccccccc}
\hline
$\qsq$ [GeV$^2$] & \multicolumn{7}{c}{$dN_{\rm corr}/d\qsq$}\\
\hline
0.80 & 0.0103 & 0.0052 & -0.0032 & -0.0033 & -0.0009 & 0.0004 & 0.0005 \\
2.38 & 0.0052 & 0.0100 & 0.0044 & 0.0011 & -0.0002 & -0.0006  & 0.0002 \\
3.97 & -0.0032 & 0.0044 &  0.0090 & 0.0048 & 0.0004 & -0.0013 & -0.0007 \\
5.56 & -0.0035 & -0.0011 &  0.0048 & 0.0070 & 0.0031 & -0.0006  & -0.0013 \\
7.14 & -0.0009 & -0.0019 &  -0.0004 & 0.0031 & 0.0044 & 0.0015 & 0.0006 \\
8.73 & 0.0004 & -0.0006 &  -0.0013 & -0.0006 & 0.0015 & 0.0023 & 0.0013 \\
10.32 & 0.0005 &  0.0002 & -0.0007 & -0.0013 & -0.0006 & 0.0013 &  0.0018 \\
\hline
%$\Lc\mun\neumb X$ d $(2.74\pm 0.02)\times 10^6$ & 
\end{tabular}}
\end{center}
\end{table}

%%\begin{table}[h]
%\begin{center}
%\caption{Covariance matrix of the measured normalized yields  ${\rm Cov}[dN_{\rm corr}(\Lb \to \Lc\mun\neumb)/d\qsq$] . }\label{tab:covmat}
%{\small
%\begin{tabular}{ccccccc}
%\hline
%0.80 & 2.38 & 3.97 & 5.56 & 7.14 & 8.73 & 10.32 \\
%\hline
%0.0103 & 0.0052 & -0.0032 & -0.0033 & -0.0009 & 0.0004 & 0.0005 \\
%0.0052 & 0.0100 & 0.0044 & 0.0011 & -0.0002 & -0.0006  & 0.0002 \\
%-0.0032 & 0.0044 &  0.0090 & 0.0048 & 0.0004 & -0.0013 & -0.0007 \\
%-0.0035 & -0.0011 &  0.0048 & 0.0070 & 0.0031 & -0.0006  & -0.0013 \\
%-0.0009 & -0.0019 &  -0.0004 & 0.0031 & 0.0044 & 0.0015 & 0.0006 \\
%0.0004 & -0.0006 &  -0.0013 & -0.0006 & 0.0015 & 0.0023 & 0.0013 \\
%0.0005 &  0.0002 & -0.0007 & -0.0013 & -0.0006 & 0.0013 &  0.0018 \\
%\hline
%%$\Lc\mun\neumb X$ d $(2.74\pm 0.02)\times 10^6$ & 
%\end{tabular}}
%\end{center}
%\end{table}

%\end{appendix}

%\end{document}
%\addcontentsline{toc}{section}{References}

%\setboolean{inbibliography}{true}
%\bibliographystyle{LHCb}
%\setboolean{inbibliography}{true}
%\bibliographystyle{LHCb}
%\bibliography{main,LHCb-PAPER,LHCb-CONF,LHCb-DP,LHCb-TDR,lb-ff}

\ifx\mcitethebibliography\mciteundefinedmacro
\PackageError{LHCb.bst}{mciteplus.sty has not been loaded}
{This bibstyle requires the use of the mciteplus package.}\fi
\providecommand{\href}[2]{#2}

\newpage

LHCb collaboration: 
R. Aaij,
B. Adeva,
M. Adinolfi,
Z. Ajaltouni,
S. Akar,
J. Albrecht,
F. Alessio,
M. Alexander,
A. Alfonso Albero,
S. Ali
% ...
\end{document}